\documentclass[conference]{IEEEtran}
\usepackage{fancyhdr}

\def\BibTeX{{\rm B\kern-.05em{\sc i\kern-.025em b}\kern-.08em
    T\kern-.1667em\lower.7ex\hbox{E}\kern-.125emX}}

\usepackage{cite}
\usepackage{xcolor}
\usepackage{subfig}
\usepackage{xcolor}
\usepackage{amsmath}
\usepackage{amssymb}
\usepackage{balance}
\usepackage{environ}
\usepackage{physics}
\usepackage{amsfonts}
\usepackage{colortbl}
\usepackage{enumitem}
\usepackage{amsfonts}
\usepackage{booktabs}
\usepackage{graphicx}
\usepackage{textcomp}
\usepackage{algorithmic}
\usepackage[normalem]{ulem}

\usepackage{tcolorbox,tikz}
\tcbuselibrary{skins}
\tcbuselibrary{breakable}

\newtcolorbox{mybox}[3][]
{
  breakable, 
  enhanced,
  colback  = #2!5, 
  colframe = #2!5,
  boxsep   = -0.5mm,
  borderline west = {1.5mm}{0.05mm}{#3!30}, 
  #1,
}

\usepackage{color}
\definecolor{blue_color}{rgb}{0.0, 0.0, 0.0}
\definecolor{red_color}{rgb}{1.0, 0.0, 0.0}

\newcommand{\sol}{\textsc{Charter}}
\newcommand{\rz}{\texttt{RZ}}
\newcommand{\x}{\texttt{X}}
\newcommand{\sx}{\texttt{SX}}
\newcommand{\cx}{\texttt{CX}}

\begin{document}

\title{\sol: Identifying the Most-Critical\\ Gate Operations in Quantum Circuits\\ via Amplified Gate Reversibility}







\author{
\IEEEauthorblockN{Tirthak Patel}
\IEEEauthorblockA{\textit{Northeastern University}\\Boston, MA, USA\\patel.ti@northeastern.edu}
\and
\IEEEauthorblockN{Daniel Silver}
\IEEEauthorblockA{\textit{Northeastern University}\\Boston, MA, USA\\silver.da@northeastern.edu}
\and
\IEEEauthorblockN{Devesh Tiwari}
\IEEEauthorblockA{\textit{Northeastern University}\\Boston, MA, USA\\d.tiwari@northeastern.edu}
}

\maketitle

\thispagestyle{fancy}
\lhead{}
\rhead{}
\chead{}
\lfoot{Author's copy (but also published in SC'22)} \rfoot{}
\cfoot{}
\renewcommand{\headrulewidth}{0pt} \renewcommand{\footrulewidth}{0pt}

\begin{abstract}

When quantum programs are executed on noisy intermediate-scale quantum (NISQ) computers, they experience hardware noise; consequently, the program outputs are often erroneous. To mitigate the adverse effects of hardware noise, it is necessary to understand the effect of hardware noise on the program output and more fundamentally, understand the impact of hardware noise on specific regions within a quantum program. Identifying and optimizing regions that are more noise-sensitive is the key to expanding the capabilities of NISQ computers.

Toward achieving that goal, we propose \sol{}, a novel technique to pinpoint specific gates and regions within a quantum program that are the most affected by the hardware noise and that have the highest impact on the program output. Using \sol{}'s methodology, programmers can obtain a precise understanding of how different components of their code affect the output and optimize those components without the need for non-scalable quantum simulation on classical computers.


\end{abstract}


\section{Introduction to \sol{}}
\label{sec:intro}

The field of quantum computing, which has applications in many areas, such as high-performance computing (HPC), has seen considerable advancement recently. While the hardware has been increasing in size, individual qubits still remain quite noisy. \textit{These noise effects include errors that are generated due to qubit state decoherence, qubit state preparation and measurement, gate operation, and cross-talk among neighboring qubits}~\cite{preskill2018quantum,patel2020experimental}. Due to these noise effects, when a quantum program is executed on these noisy intermediate-scale quantum (NISQ) computers, the program output experiences a considerable error~\cite{preskill2018quantum,patel2020experimental,liu2020reliability,butko2019understanding,wright2019benchmarking}.

As practical hardware error correction has not been widely implemented on NISQ computers in a cost-effective manner~\cite{vuillot2019quantum,terhal2019scalable,sun2019experimental,layden2019ancilla}, recent software efforts have attempted to reduce the output error using software-level compiler optimizations~\cite{li2022paulihedral,bhattacharjee2019muqut,tannu2019ensemble,davis2020towards}. These software efforts rely on relatively simple noise models -- for example, individual-qubit noise data generated during qubit calibration~\cite{tannu2019not,patel2020veritas} -- to obtain the decoherence times and error rates of gate operations on individual qubits. This data is then used to perform just-in-time compilation to best map the logical quantum program to the physical quantum computer in a manner that reduces the number of gate operations and potentially reduces the overall error in the output of the program~\cite{wilson2020just,patel2020ureqa,tannu2019not,murali2019noise,patel2020veritas}.

\begin{figure}[t]
    \centering
    \subfloat[Location of Two-Qubit Gates]{\includegraphics[scale=0.45]{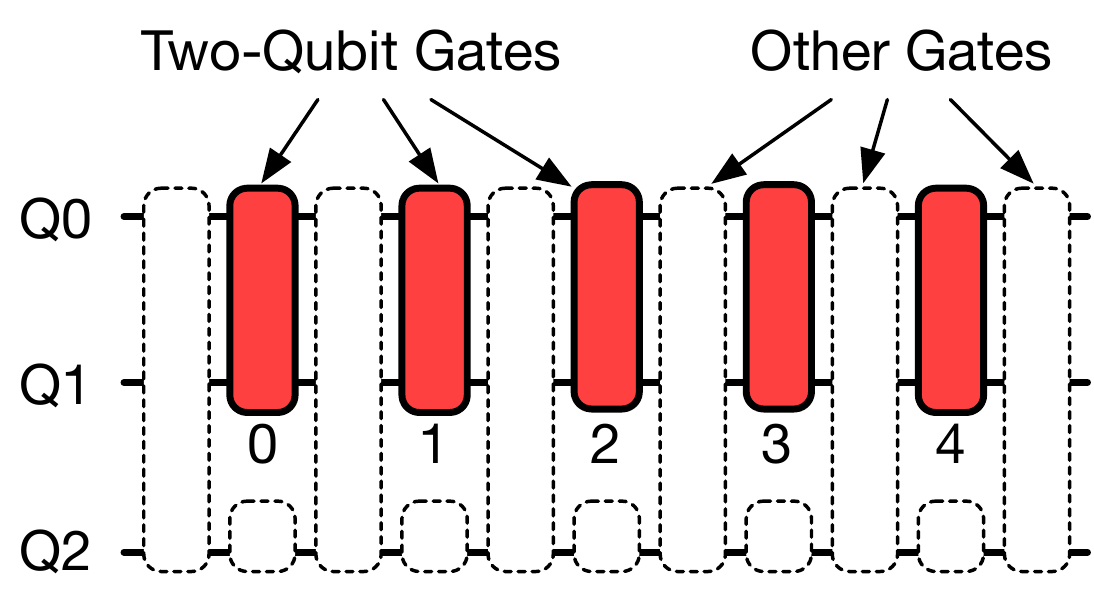}}\hfill
    \subfloat[Error of Each Gate]{\includegraphics[scale=0.51]{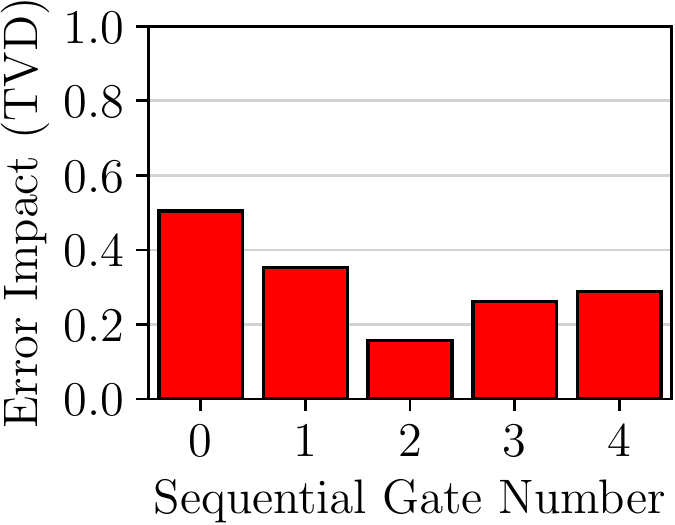}}
    \vspace{1mm}
    \hrule
    \vspace{1mm}
    \caption{(a) The same two-qubit gate run on the same qubits for the Quantum Fourier Transform (QFT) program. (b) The gate has a different error impact on the program output each time it is run (expressed as Total Variation Distance (TVD) on the Y-axis -- refer to Sec.~\ref{sec:meth} for more methodological details).}
    \label{fig:motiv}
    \vspace{-3mm}
\end{figure}

\textit{However, due to the difficult-to-model complex interactions of the different noise characteristics, how different quantum program components contribute to the overall output error of a quantum program can vary from program to program and program region to program region}~\cite{patel2020experimental,liu2020reliability,butko2019understanding,wright2019benchmarking}. Fig.~\ref{fig:motiv} shows the error impact of a two-qubit gate being run on the same pair of physical qubits at five different times during the program execution. These results show that the magnitude of impact on the program output of the same gate is quite different depending on its position in the program (i.e., the TVD changes depending on the location of the erroneous two-qubit gate operation). The fundamental reason behind our observation is that the crosstalk and interference from gates being applied to neighboring qubits affects a particular gate operation's impact on the error in the program output.\textit{ Unfortunately, quantifying, characterizing and modeling this complex effect is challenging, and currently, the HPC quantum systems community does not have the appropriate methodology and tools to determine how the specific components/regions of a quantum program contribute to its output error.} Novel methods and tools in this domain will enable further program-specific compiler optimizations and debugging support. 


\vspace{3mm}

\noindent\textbf{\sol{}:} In order to address the above challenges, this paper proposes \sol{}, a simple yet effective method to compute how specific gates within a program contribute to its overall output error. \sol{} enables the programmer to pinpoint specific gate operations and regions within a quantum program that are the most affected by the hardware noise and that have the highest impact on the output error of the program. The spirit of this effort is similar to development of program vulnerability factor (PVF) and architecture vulnerability factor (AVF) for classical computing systems~\cite{sridharan2009eliminating,mukherjee2003systematic} -- however, developing similar closed-form models for highly accurate vulnerability assessment for quantum circuits is not possible.




The key insight behind \sol{} is to leverage the reversibility property of quantum computing to understand the impact of individual quantum gate operations on the observed error in the program output. \sol{} is the first method to leverage the quantum circuit reversal to identify and quantify the criticality of individual quantum gate operations on the program output error (Sec.~\ref{sec:desi}). \sol{} demonstrates how to systematically reveal the impact of individual quantum gate operations \textit{without the need to perform quantum simulation on classical computers, which is not scalable to large programs~\cite{cirac2012goals,khammassi2017qx,jones2019quest,li2019sanq}}. \sol{} breaks this barrier and demonstrates how quantum programmers can identify the most critical operations on erroneous NISQ hardware without relying on expensive simulation of quantum programs. \textit{Using \sol{}'s methodology, programmers can identify particularly problematic program regions and focus their debugging and optimization efforts on those areas.}

A prior technique, \textsc{Qraft}~\cite{patel2021qraft}, demonstrated the use of the quantum circuit reversibility toward improving the answer fidelity of quantum programs on the NISQ computers (i.e., circuit reversal as a noise-mitigation technique). This work, \sol{}, demonstrates another unique and useful application of quantum circuit reversibility -- estimating the high-impact quantum circuit components. \textsc{Qraft} reverses the full quantum circuit to mitigate the noise and estimate correct program output~\cite{patel2021qraft}. Unlike, \textsc{Qraft}, \sol{} considers the quantum circuit as a series of smaller circuit gate operations, and applies quantum circuit reversal at the gate-level. \sol{} demonstrates the significance of applying circuit reversals multiple times for a given circuit gate and across multiple circuit gates -- which is necessary to identify the high-impact program components. 


\vspace{3mm}

\noindent\textbf{The contributions of this work are as follows:}
\begin{itemize}[leftmargin=*]
    \item The proposed technique, \sol{}, is a simple and effective way to determine high-error-impact gate operations and regions in any given quantum program using quantum reversibility property. \textit{It gives programmers a scalable utility to perform optimizations that are specific to their program structure, as opposed to relying on simplistic assumptions and noise models.}
    
    \vspace{1mm}
    
    \item \sol{} is evaluated using multiple quantum programs of varying characteristics and sizes. \textit{These results have revealed some surprising, interesting, and useful trends:}
    
    \vspace{1mm}
    
    \begin{itemize}
        
        \item[+] While most existing works have primarily focused on ranking the criticality of qubits based on their error rates~\cite{wilson2020just,patel2020ureqa,tannu2019not,murali2019noise,patel2020veritas}, \textit{\sol{} demonstrates that prior approaches alone are not always sufficient and complete. Type and location of gate operations are equally critical. Across different algorithms, the top 25\% high-impact gates are located on almost all qubits used to run the program, which indicates that criticality of gates varies more based on the gate's position in the circuit.}
        
        \vspace{1mm}
    
        \item[+] Prior techniques have largely focused on specifically reducing two-qubit gates~\cite{smith2019quantum,zulehner2019compiling,wille2019mapping,ash2019qure,patel2022quest}, as they generally have a high error in isolation. \textit{Interestingly, \sol{} reveals that, in some cases, one-qubit gates can have worse impacts on the output error.} 
        
        \vspace{1mm}
        
        \item[+] \textit{\sol{} demonstrates that the error impact of specific gates can vary depending on the input to the program.} For example, input $i$ can increase the error impact of gates in a certain region of a quantum program, while input $j$ can affect a different region. This observation opens up the opportunity for input-aware quantum compilation. 
        
        \vspace{1mm}
        
        \item[+] \textit{We provide a technique to identify the program input with the highest impact on the output error by extending \sol{} to multi-gate reversals. Based on this identification, we propose the use of selective serialization of high-impact gates to reduce the effect of crosstalk, which reduces output error by up to 7\% points.}
        
        \vspace{1mm}
        
        \item[+] \textit{\sol{} shows that the gates near the end of the program do not always have the worst impact}, although some existing techniques optimize those regions due to the assumed impact of qubit state decoherence~\cite{liu2021relaxed,shi2019optimized}, which increases as program length increases~\cite{mavadia2017prediction,PhysRevA.58.2733}. 
        
    \end{itemize}
    
    \vspace{1mm}
    
    \item \textit{\sol{}'s code to apply reversals and use them for characterization, and its evaluation datasets are  open-sourced at} \texttt{https://doi.org/10.5281/zenodo.6875324}.
\end{itemize}

\section{Relevant Background for \sol{}}
\label{sec:motiv}

\begin{figure}[t]
    \centering
    \subfloat[Quantum Gates]{\includegraphics[scale=0.4]{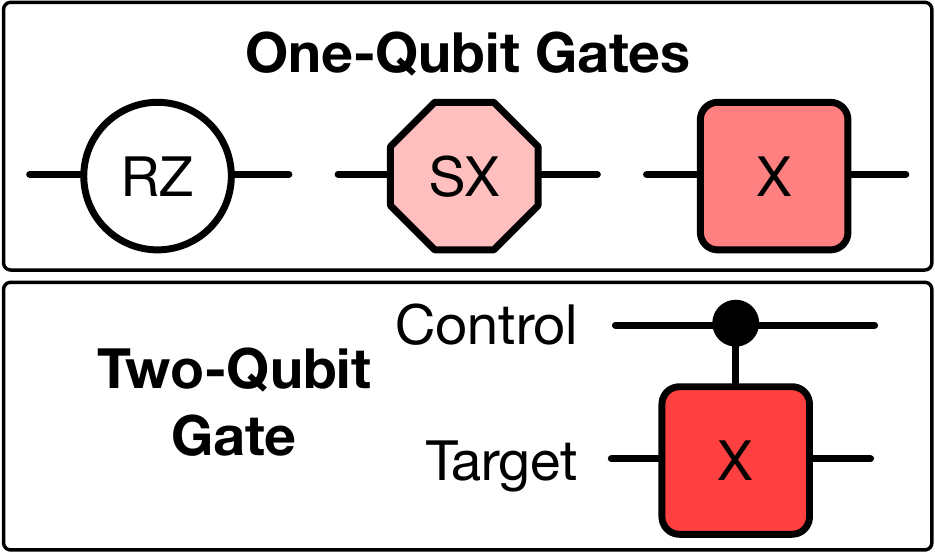}}\hfill
    \subfloat[Quantum Circuit]{\includegraphics[scale=0.4]{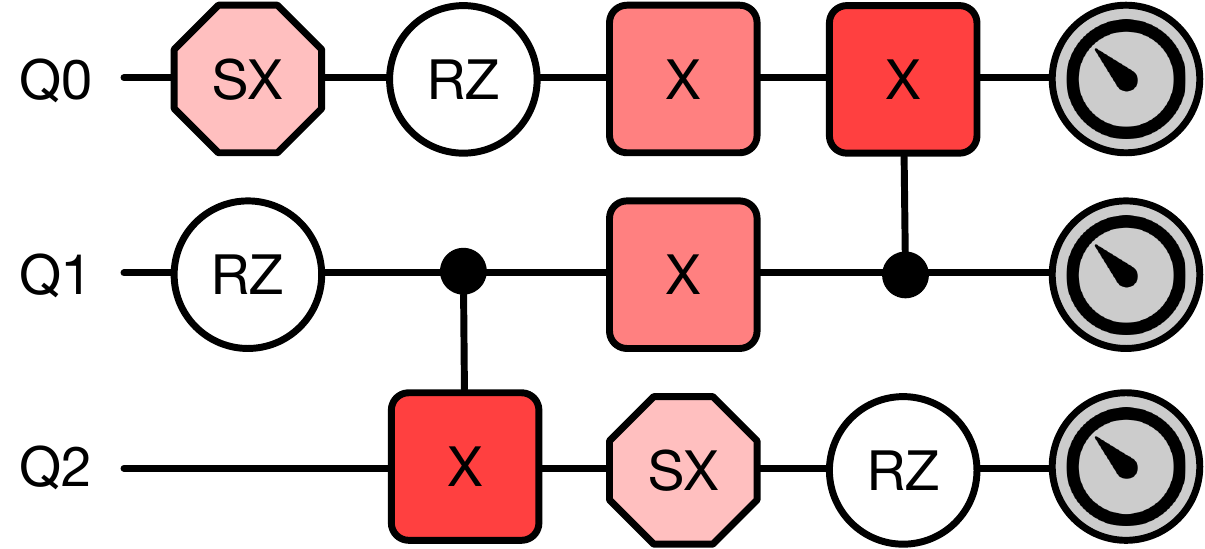}}
    \vspace{1mm}
    \hrule
    \vspace{1mm}
    \caption{(a) A set of one- and two-qubit basis gates is used to perform universal quantum computation on IBM computers. (b) An example three-qubit quantum circuit with gates applied from left to right. The qubit state evolution is represented using horizontal lines; at the end of the quantum computation, the measurement operations are applied to read the qubit state.}
    \label{fig:basis}
    \vspace{-3mm}
\end{figure}

\noindent\textbf{Quantum States, Gates, and Circuits.} The computational state of a qubit can be represented using a \textit{superposition} of the $\ket{0}$ and $\ket{1}$ basis states: $\ket{\psi} = \alpha_1\ket{0} + \alpha_2\ket{1}$. Here, $\ket{\psi}$ is the superposed state of the qubit and $\alpha_1$ and $\alpha_2$ are complex coefficients such that $\norm{\alpha_1}^2 + \norm{\alpha_2}^2 = 1$. Similarly, the superposed and \textit{entangled} state of an $n$-qubit quantum system can be represented as: $\ket{\psi} = \sum_{k=0}^{k=2^n-1}\alpha_k\ket{k}$ such that $\sum_{k=0}^{k=2^n-1}\norm{\alpha_k}^2 = 1$. When the qubit state is measured, the state superposition collapses and it can be found in one of the $2^n$ states, with the probability of finding it in state $k$ being $\norm{\alpha_k}^2$.

Qubit states are prepared by applying \textit{quantum gates or operations}. Quantum gates are unitary operations that evolve the state of the qubit from one state to another. For example, the \x{} quantum gate flips the state of the qubit from $\ket{0}$ to $\ket{1}$ and from $\ket{1}$ to $\ket{0}$. A set of \textit{universal quantum basis gates} can be used to represent all quantum gates. The specific choice of basis gates depends on the hardware. For example, the IBM quantum computing hardware, which is used to evaluate \sol{}, uses the \rz{}, \sx{}, \x{}, and \cx{} gates to form the universal basis set. As Fig.~\ref{fig:basis}(a) shows, \rz{}, \sx{}, and \x{} are one-qubit gates, and \cx{} is a two-qubit gate. The \x{} gate is the qubit-flip gate, the \sx{} gate performs the square root operation of the \x{} gate, and the \rz{} gate is a virtual gate that changes the frame of reference of the quantum system (it is not performed on the physical hardware). The two-qubit entangling \cx{} gate performs the \x{} operation on the ``target'' qubit depending on the state of the ``control'' qubit as shown in Fig.~\ref{fig:basis}(a).

When these quantum gates are performed in succession of one another on an $n$-qubit system they form a quantum program, also known as a quantum circuit. While a quantum program can be written using any high-level quantum logic gates, the program then has to be mapped to the hardware-compatible physical basis gate set in order to execute the circuit on a quantum computer. Fig. ~\ref{fig:basis}(b) shows an example three-qubit quantum circuit with nine operations. Once all the operations are executed, the measurement operator is applied to all the qubits to generate the output state.

\begin{table}[t]
    \centering
    \caption{Different types of noise effects that NISQ computers experience when running a quantum circuit.}
    \vspace{-1mm}
    \begin{tabular}{|>{\columncolor[gray]{0.85}}p{27mm}|p{50mm}|}
        \hline
        \textbf{Noise Type} & \textbf{Description} \\
        \hline
        \hline
        \textit{Operation Errors} \newline \cite{erhard2019characterizing,proctor2019direct} & Errors caused by inaccuracies in applying a specific desired quantum gate.\\
        \hline
        \textit{State Preparation and Measurement Errors} \newline \cite{tannu2019mitigating} & Errors caused when initializing and preparing the qubit state and measuring the final state after execution.\\
        \hline
        \textit{Decoherence Effects} \newline \cite{mavadia2017prediction,PhysRevA.58.2733} & Errors caused due to the prepared qubit superposition state decaying over time to the ground (i.e., $\ket{0}$) state.\\
        \hline
        \textit{Crosstalk Effects} \newline \cite{murali2020software,xie2022suppressing,xie2021mitigating} & Errors caused to a qubit due to interference effects from neighboring qubits.\\
        \hline
    \end{tabular}
    \label{tab:noise}
    \vspace{0mm}
\end{table}

\begin{figure}[t]
    \centering
    \includegraphics[scale=0.42]{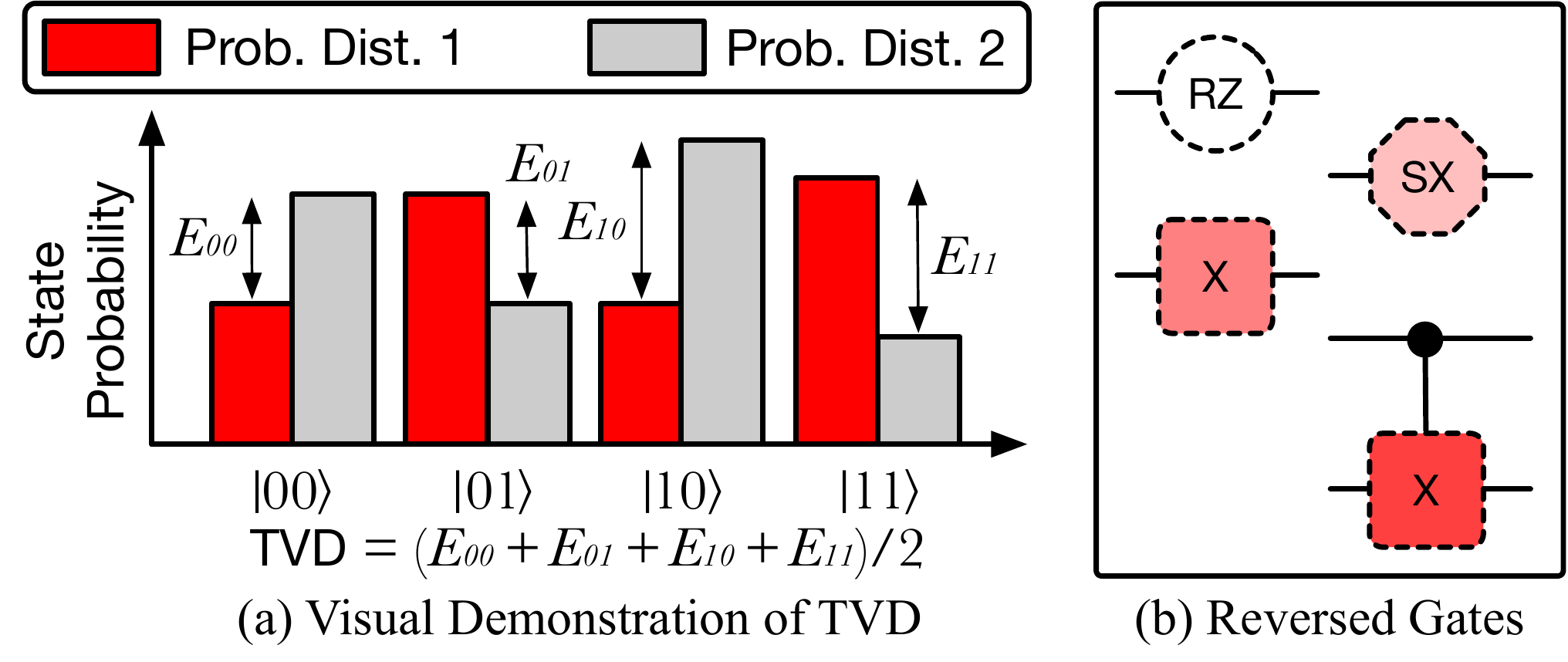}
    \vspace{1mm}
    \hrule
    \vspace{1mm}
    \caption{(a) The Total Variation Distance (TVD) between two probability distributions is measured by summing the differences in state probabilities of the two distributions and dividing by two. (b) The reverse versions of the basis gates are depicted with dashed borders in this paper.}
    \label{fig:tvd}
    \vspace{-3mm}
\end{figure}

\vspace{3mm}

\noindent\textbf{Hardware Noise Effects.} A NISQ computer experiences a variety of noise effects when a quantum program is initialized, executed, and measured. These noise effects have been summarized in Table~\ref{tab:noise}. \textit{These noise effects are challenging to model; especially the impact of their combined expression is difficult to simulate or predict}~\cite{erhard2019characterizing,preskill2018quantum,butko2019understanding}. When a quantum circuit is run, these individual noise effects compound and accumulate to cause error in the output of the program.

\vspace{3mm}

\noindent\textbf{Quantum Output and Output Error.} The output of a quantum program is obtained as a probability distribution over its output states. The same quantum state is prepared and measured multiple times (each measurement producing one output state), which then generates the probability distribution over the output states. However, because of the hardware noise effects, this observed output distribution can be quite different from the ideal or correct output distribution. A typical metric that is used to measure the deviation of the observed output distribution from the ideal output distribution is the \textit{Total Variation Distance or TVD}~\cite{gilchrist2005distance}. As shown in Fig.~\ref{fig:tvd}(a), the TVD between two probability distributions is calculated by summing up the absolute differences in the probabilities of the individual states and dividing by two. With a value between 0 and 1, the closer the TVD is to 0, the more the two probability distributions are similar. We will use this metric going forward to compare two probability distributions.

\vspace{3mm}

\noindent\textbf{Quantum Reversibility Property.} Unlike classical gates, all quantum gates are reversible. The input state can be obtained by applying the inverse of the original gate to the output state. Because all quantum gates can be represented using unitary matrices $U$, their inverse is simply the complex conjugate transpose (also known as the Hermitian adjoint) of $U$, which can be represented as $U^\dagger$. Therefore, for any initial state $\ket{\psi}$, if $U$ is applied to it, the initial state can be recovered by applying $U^\dagger$ ($\ket{\psi} = U^\dagger U\ket{\psi}$). In this paper, we represent the inverse of a quantum operation using a dashed line as shown in Fig.~\ref{fig:tvd}(b). QRAFT has leveraged this property for improving the answer fidelity of quantum programs on NISQ machines~\cite{patel2021qraft}. \sol{} leverages this quantum reversibility property to identify high-impact circuit components.
\section{\sol{}'s Methodology for Experiments}
\label{sec:meth}

\begin{figure}[t]
    \centering
    \includegraphics[scale=0.29]{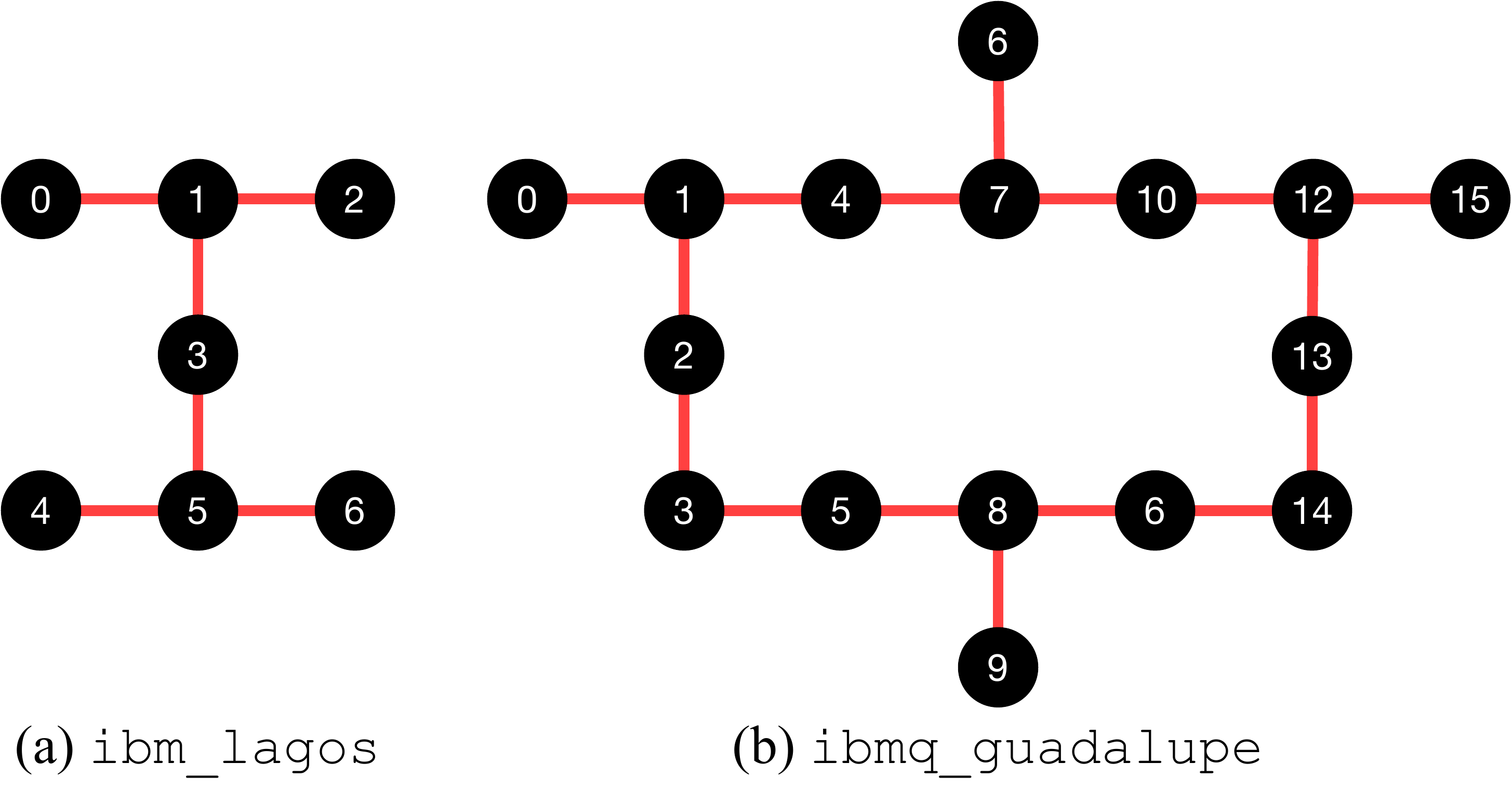}
    \vspace{1mm}
    \hrule
    \vspace{1mm}
    \caption{Qubit connectivity layout of quantum computers (a) \texttt{ibm\_lagos} and (b) \texttt{ibmq\_guadalupe}.}
    \label{fig:comps}
    \vspace{-1mm}
\end{figure}

\begin{table}[t]
    \centering
    \caption{Algorithms and benchmarks used for the design, evaluation, and analysis of \sol{}.}
    \vspace{-1mm}
    \begin{tabular}{|>{\columncolor[gray]{0.85}}p{15mm}|p{62mm}|}
        \hline
        \textbf{Algorithm} & \textbf{Description} \\
        \hline
        \hline
        HLF & Hidden Linear Function~\cite{bravyi2018quantum} \\
        \hline
        QFT & Quantum Fourier Transform~\cite{namias1980fractional} \\
        \hline
        QAOA & Quantum Alternating Operator Ansatz~\cite{farhi2016quantum} \\
        \hline
        VQE & Variational Quantum Eigensolver~\cite{mcclean2016theory} \\
        \hline
        Adder & Quantum Adder Program~\cite{cuccaro2004new} \\
        \hline
        Multiplier & Quantum Multiplier Program~\cite{hancock2019cirq} \\
        \hline
        Heisenberg & Time-independent Heisenberg Hamiltonian~\cite{bassman2021arqtic} \\
        \hline
        TFIM & Transverse Field Ising Model~\cite{bassman2021arqtic}  \\
        \hline
        XY & XY Quantum Heisenberg Model~\cite{bassman2021arqtic} \\
        \hline
    \end{tabular}
    \label{tab:algorithms}
    \vspace{0mm}
\end{table}

\begin{figure*}[t]
    \centering
    \subfloat[\sx{} Gate Reversal]{\includegraphics[scale=0.33]{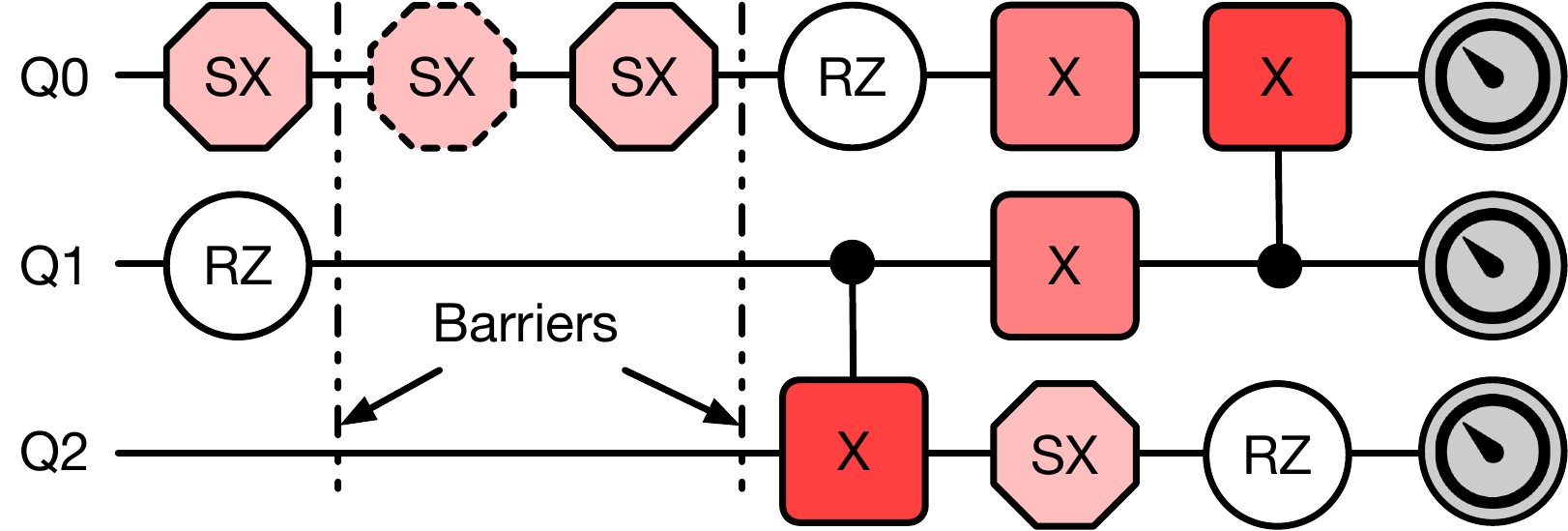}}\hfill
    \subfloat[\cx{} Gate Reversal]{\includegraphics[scale=0.33]{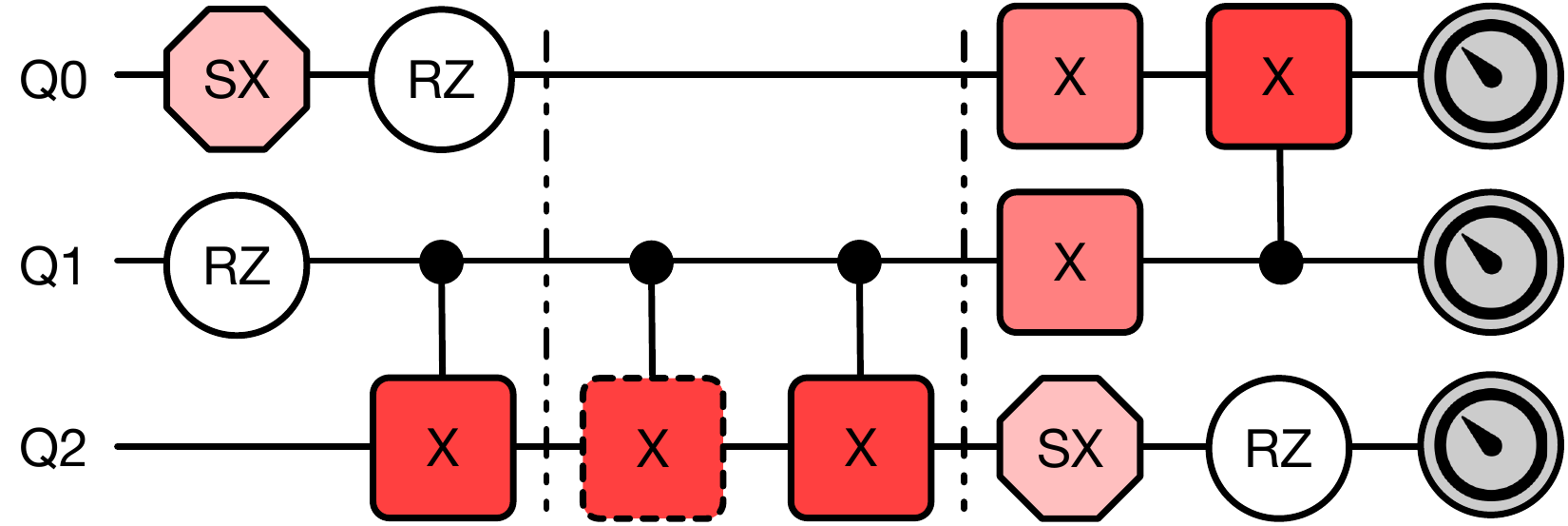}}\hfill
    \subfloat[Two \sx{} Gate Reversals]{\includegraphics[scale=0.33]{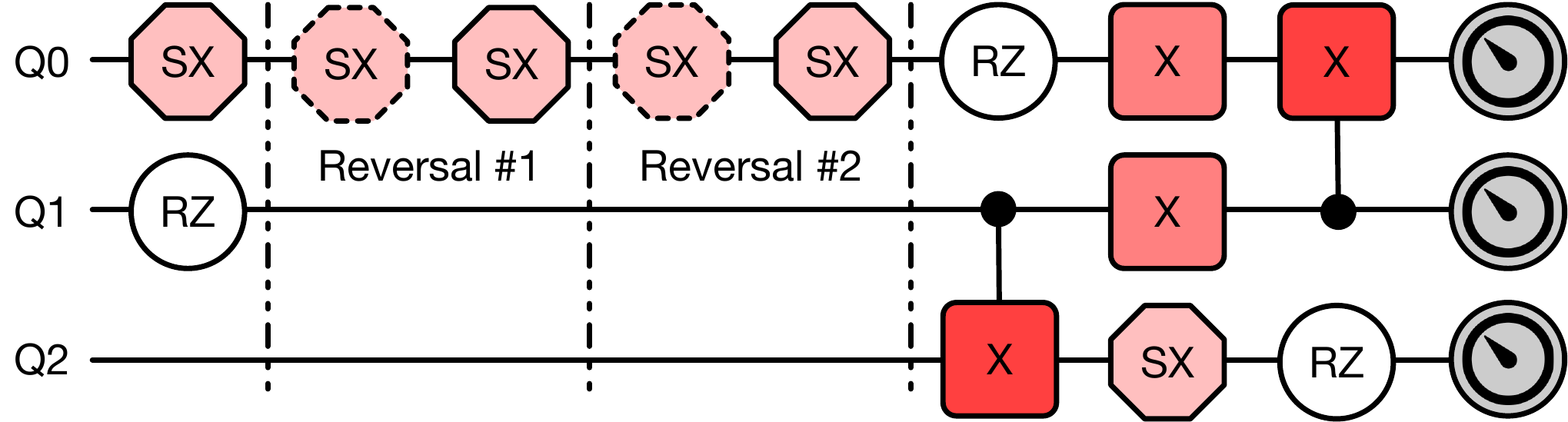}}
    \vspace{1mm}
    \hrule
    \vspace{1mm}
    \caption{Example of applying a reversed pair to one of the (a) \sx{} and (b) \cx{} gates. (c) Example of applying two reversals to the \sx{} gate. ``Barriers'' are used to ensure that other operations are not run in parallel with the reversed pair in order to isolate for the error impact of the gate in question (other qubits are kelp idle when the reversed pair is being executed). }
    \label{fig:sx_cx}
    \vspace{-3mm}
\end{figure*}

Before we describe the design of \sol{}, we first discuss the methodology to construct and execute the experiments used for the design, evaluation, and analysis of \sol{}.

\vspace{3mm}

\noindent\textbf{Experimental Setup.} We use the IBM quantum cloud platform~\cite{castelvecchi2017ibm} to execute the experiments to evaluate \sol{}. We use the corresponding Python-based (Python version 3.9.7) Qiskit programming language~\cite{qiskit} (Qiskit version 0.17.4) to program the quantum algorithms and benchmarks and apply reversals for \sol{}. We run the experiments on the 7-qubit \texttt{ibm\_lagos} quantum computer and the 16-qubit \texttt{ibmq\_guadalupe} quantum computer (computer qubit layouts shown in Fig.~\ref{fig:comps}), depending on the size of the algorithm being executed; algorithms of size seven qubits or less are run on \texttt{ibm\_lagos} and the rest are run on \texttt{ibmq\_guadalupe}. The quantum volume of both computers is 32; this metric measures the performance of gate-model quantum computers in terms of the size (width and depth) of the circuits they can run. Each experiment is run with 32,000 trials to generate the output probability. We use Qiskit's StateVector simulator to run simulations of the quantum programs on classical computers to get the ideal output of the programs. \textit{Note that this output is only used to validate the effectiveness of \sol{} and is not a part of the \sol{} technique.} In fact, \sol{} is specifically designed to not rely on classical simulations in order to ensure scalability to large quantum algorithms. Python's SciPy package~\cite{jones2016scipy} is used to calculate statistical metrics such as correlation for the design and analysis of \sol{}. 

\vspace{3mm}

\noindent\textbf{Quantum Algorithms and Benchmarks.} \sol{} uses algorithms and benchmarks listed in Table~\ref{tab:algorithms}, which have a variety of sizes (3-16 qubits) and use cases for high-performance computing. HLF is a search algorithm, while QFT is the quantum version of the Fourier transform. QAOA and VQE are quantum variational algorithms, and Adder and Multiplier are standard quantum arithmetic circuits. Heisenberg, TFIM, and XY are time-evolution Hamiltonian programs for material simulations. The TFIM model has non-zero coupling interaction between nearest neighbor spins along the $z$ axis, XY has for $x$ and $y$ axes, and Heisenberg has for all three axes. \textit{All algorithms are mapped to the physical quantum hardware using all of the compiler optimizations available via Qiskit, including noise-aware logical-to-physical mapping, gate routing and SWAP insertion depending on hardware connectivity, gate cancellation based on commutativity and combination, gate deletion by detecting redundancy, and gate re-synthesis for two-qubit unitary blocks (peep-hole optimization)~\cite{qiskit}.} All algorithms are converted to the \{\cx{}, \rz{}, \sx{}, \x{}\} basis gate set (Fig.~\ref{fig:basis}) during the mapping process to run on the IBM hardware. Once \sol{}'s techniques are applied to the circuit, the compiler optimization level is set to 0 to ensure that the techniques are not optimized out, because \sol{} is applied to pre-mapped circuit and no additional compilation steps necessary.

\vspace{3mm}

\noindent\textbf{Metrics.} We use the TVD (described in the previous section: Sec.~\ref{sec:motiv}) to measure the difference between two probability distributions. To measure the correlation relationship between two random variables, we use the Pearson correlation~\cite{benesty2009pearson}, which is the square-root of the coefficient of linear regression ($R^2$). A value of 1 indicates a perfect positive linear correlation, $-1$ indicates a perfect negative linear correlation, and 0 indicates no correlation. We also show the $p$-value to show the significance of the correlation (a $p$-value $<0.01$ indicates a statistically strong correlation).

\section{\sol{}: Design and Technique}
\label{sec:desi}

\begin{table*}[t]
    \centering
    \caption{Pearson correlation between the TVD of the output of the $r$-reversal runs with the output of the ideal simulation and the TVD of the output of the $r$-reversals runs with the output of the original run for different number of reversals ($r$). The number in brackets next to the algorithm name shows the number of qubits. The highest correlation is seen with five reversals.}
    \vspace{-1mm}
    \begin{tabular}{|>{\columncolor[gray]{0.85}}p{20mm}|p{17mm}|p{13mm}|p{17mm}|p{13mm}|>{\columncolor[RGB]{255, 200, 200}}p{17mm}|>{\columncolor[RGB]{255, 200, 200}}p{13mm}|p{17mm}|p{13mm}|}
        \hline
        & \multicolumn{2}{|c|}{\textbf{One Reversal}} & \multicolumn{2}{|c|}{\textbf{Three Reversals}} & \multicolumn{2}{|>{\columncolor[RGB]{255, 200, 200}}c|}{\textbf{Five Reversals}} & \multicolumn{2}{|c|}{\textbf{Seven Reversals}} \\
        \hline
        \textbf{Algorithm} & \textbf{Correlation} & \textbf{$p$-value} & \textbf{Correlation} & \textbf{$p$-value} & \textbf{Correlation} & \textbf{$p$-value} & \textbf{Correlation} & \textbf{$p$-value} \\
        \hline
        \hline
        HLF (5) & 0.02 & 0.91 & 0.08 & 0.58 & 0.40 & 6.85$e-$3 & 0.17 & 0.26 \\
        \hline
        HLF (10) & 0.11 & 0.06 & 0.18 & 0.12 & 0.49 & 3.04$e-$3 & 0.13 & 0.62 \\
        \hline
        QFT (3) & 0.43 & 3.69$e-$3 & 0.96 & 3.78$e-$24 & 0.99 & 2.18$e-$35 & 0.99 & 2.93$e-$40 \\
        \hline
        QFT (7) & 0.61 & 3.18$e-$31 & 0.61 & 9.21$e-$31 & 0.64 & 3.32$e-$34 & 0.63 & 2.93$e-$34 \\
        \hline
        Adder (4) & 0.52 & 3.29$e-$7 & 0.94 & 9.57$e-$41 & 0.98 & 3.97$e-$61 & 0.99 & 5.40$e-$68 \\
        \hline
        Adder (9) & 0.43 & 5.12$e-$56 & 0.89 & 5.89$e-$23 & 0.94 & 4.12$e-$43 & 0.95 & 2.31$e-$30 \\
        \hline
        Multiply (5) & 0.76 & 2.98$e-$17 & 0.96 & 1.32$e-$48 & 0.99 & 1.04$e-$67 & 0.99 & 7.85$e-$75 \\
        \hline
        Multiply (10) & 0.89 & 7.35$e-$221 & 0.89 & 1.45$e-$219 & 0.89 & 3.16$e-$220 & 0.88 & 7.20$e-$217 \\
        \hline
        QAOA (5) & 0.82 & 4.34$e-$35 & 0.70 & 1.13$e-$21 & 0.79 & 6.02$e-$31 & 0.80 & 1.80$e-$31 \\
        \hline
        QAOA (10) & 0.38 & 6.65$e-$15 & 0.35 & 4.21$e-$13 & 0.38 & 2.46$e-$10 & 0.30 & 1.11$e-$9 \\
        \hline
        VQE (4) & 0.51 & 1.69$e-$29 & 0.38 & 4.11$e-$16 & 0.21 & 9.97$e-$6 & 0.19 & 6.83$e-$5 \\
        \hline
        Heisenberg (4) & 0.69 & 2.29$e-$28 & 0.74 & 2.27$e-$29 & 0.90 & 4.75$e-$36 & 0.91 & 4.50$e-$47 \\
        \hline
        TFIM (4) & 0.70 & 1.09$e-$18 & 0.78 & 1.67$e-$25 & 0.88 & 4.70$e-$38 & 0.92 & 1.50$e-$49 \\
        \hline
        TFIM (8) & 0.38 & 1.37$e-$20 & 0.53 & 7.67$e-$16 & 0.71 & 7.05$e-$14 & 0.60 & 1.07$e-$12 \\
        \hline
        TFIM (16) & 0.42 & 2.73$e-$19 & 0.55 & 4.69$e-$18 & 0.72 & 8.14$e-$16 & 0.59 & 2.61$e-$13 \\
        \hline
        XY (4) & 0.49 & 6.50$e-$7 & 0.84 & 1.54$e-$26 & 0.91 & 2.13$e-$36 & 0.92 & 6.70$e-$39 \\
        \hline
        XY (8) & 0.67 & 0.11 & 0.76 & 3.87$e-$4 & 0.80 & 9.87$e-$6 & 0.89 & 5.44$e-$11 \\
        \hline
    \end{tabular}
    \label{tab:revcorr}
    \vspace{0mm}
\end{table*}

The core of \sol{}'s design is grounded in the reversibility principle of quantum computing. \sol{} leverages this property by applying a pair consisting of the ``reversed'' gate and the ``original'' gate to each of the gates in the quantum circuit (referred to as the ``reversed pair''). Fig.~\ref{fig:sx_cx} shows how this pair is applied to different operations: (a) the first \sx{} gate on qubit 0 and (b) the first \cx{} gate on qubits 1 and 2. This is done for each of the gates in the circuit. The original circuit has nine gates; therefore, a total of nine ``reversed circuits'' are generated, each with one reversed pair of gates corresponding to one of the gates in the original circuit. In general, $g$ reversed circuit are generated for a circuit with $g$ gates.

\sol{} employs this method for two useful reasons. First, applying the pair of the reversed gate + the original gate to a gate $U$ does not alter the ground truth output of the circuit (on an ideal quantum computer). This is because of the reversible principle of unitary matrices: $U(U^\dagger U) = UI = U$. Therefore, \textit{applying this pair does not alter the mathematical functionality of the quantum program, and does not affect the output of the program}. (2) However, due to the introduction of additional two operations within the reversed pair, this will impact the output error of the circuit when it is run on a real noisy quantum computer. This impact on the output error will be the amplified, but it will be proportional to the impact of the original gate. This is because the two gates in the reversed pair are operationally similar to the original gate, meaning they will experience the same operation errors. Moreover, the two gates in the reversed pair are also applied at the same location of the original gate, which means that they will also experience the same decoherence effects. Thus, \textit{the output of the original circuit (i.e., the circuit with no reversals) and the output of the circuit with the reversed pair can be compared to gauge the impact of the gate in question}.

These outputs are compared in the following manner. Let $O_{orig}$ be the output of the original circuit. $O_{orig}$ is a probability distribution with probabilities $p^k_{orig}$ for each of the $k^{th}$ state in the $2^n$ states, where $n$ is the number of qubits. Similarly, let $O_{rev}$ be the output of the reversed circuit for a particular gate with $p^k_{rev}$ probabilities for each of the  $2^n$ states. The difference in these outputs can then be calculated using the total variation distance: $TVD\big{(}O_{orig},O_{rev}\big{)} = \frac{1}{2}\sum^{k=2^n-1}_{k=0} \Big{|}p^k_{orig}-p^k_{rev}\Big{|}$. Using this TVD, \sol{} can estimate the impact that a particular gate has on the output of a circuit. Note that \textit{this TVD is calculated between the noisy outputs of the original circuit and the reversed circuit. It does not rely on the output of an ideal quantum simulation, which is not scalable as it has to be executed on classical computers}~\cite{cirac2012goals,khammassi2017qx,jones2019quest,li2019sanq}. \textit{Instead, \sol{} solely relies on noisy outputs of real-computer runs}. To get the error impact of all the gates, the TVD can be calculated between the output of the original circuit and the reversed circuits corresponding to all the gates. Going forward, we refer to gates whose circuits have high TVD with the original circuit as ``high-impact'' gates as these are the ones of particular importance.

As a note, the TVD measures the relative error between the two output probability distributions. As a result, the absolute rates of errors unrelated to gate operations, including state preparation and measurement errors, remain the same for both of the circuits and therefore, do not affect \sol{}'s results. Note also that this methodology of CHARTER is not affected by intermediate measurements and resets. For example, to gauge the impact of a particular gate before an intermediate measurement, one can apply reversals before the measurement. A gate can similarly be reversed after the intermediate measurement and its impact will be observed in the succeeding measurement.

Next, we look at why identifying high-impact gates in noisy environments is a challenge and how \sol{} mitigates it.

\begin{table}[t]
    \centering
    \caption{The number and percentage of \rz{} and \cx{} gates in different algorithms and their corresponding depths. The percentage of \rz{} gates in a circuit is high across algorithms.}
    \vspace{-1mm}
    \begin{tabular}{|>{\columncolor[gray]{0.85}}p{17mm}|p{9mm}|>{\columncolor[RGB]{255, 200, 200}}p{9mm}|p{9mm}|p{9mm}|p{9mm}|}
        \hline
        \textbf{Algorithm} & \textbf{Num. \rz{}s} & \textbf{\% \rz{}s} & \textbf{Num. \cx{}s} & \textbf{\% \cx{}s} & \textbf{Depth}\\
        \hline
        \hline
        HLF (5) & 14 & 41\% & 10 & 29\% & 31 \\
        \hline
        HLF (10) & 62 & 22\% & 171 & 61\% & 79 \\
        \hline
        QFT (3) & 18 & 42\% & 9 & 21\% & 28 \\
        \hline
        QFT (7) & 121 & 42\% & 88 & 30\% & 141 \\
        \hline
        Adder (4) & 35 & 41\% & 24 & 28\% & 61 \\
        \hline
        Adder (9) & 99 & 28\% & 212 & 60\% & 209 \\
        \hline
        Multiply (5) & 32 & 37\% & 29 & 34\% & 58 \\
        \hline
        Multiply (10) & 206 & 31\% & 332 & 51\% & 321 \\
        \hline
        QAOA (5) & 51 & 37\% & 55 & 40\% & 84 \\
        \hline
        QAOA (10) & 107 & 26\% & 222 & 53\% & 173 \\
        \hline
        VQE (4) & 172 & 40\% & 132 & 31\% & 264 \\
        \hline
        Heisenberg (4) & 171 & 33\% & 201 & 39\% & 338 \\
        \hline
        TFIM (4) & 48 & 41\% & 33 & 28\% & 62 \\
        \hline
        TFIM (8) & 223 & 41\% & 137 & 25\% & 168 \\
        \hline
        TFIM (16) & 1032 & 36\% & 1000 & 35\% & 499 \\
        \hline
        XY (4) & 35 & 37\% & 31 & 33\% & 64 \\
        \hline
        XY (8) & 178 & 36\% & 149 & 30\% & 183 \\
        \hline
    \end{tabular}
    \label{tab:rzcxs}
    \vspace{0mm}
\end{table}

\subsection{\sol{}: Multiple Reversals for Noise Mitigation}

While designing \sol{}, we discovered that although the idea of gate-specific reversed circuits is intuitive and promising, it is often not as effective as expected in practice. This is because, due to the high noise levels on NISQ computers, applying one pair of reversals does not help us effectively identify the differences between the outputs of the two circuits with high statistical confidence -- the TVD between the original circuit and the reversed circuit may be close to zero in some cases. This is especially true for large circuits with many qubits and gates as there is a large enough accumulated impact from other gates, so as to obscure the impact of one particular gate. To overcome this challenge, \sol{} designs a novel multiple-reversal technique. \sol{} applies the reversed pair multiple times in succession of one another. Fig.~\ref{fig:sx_cx}(c) shows how ``two reversals'' can be applied to the first \sx{} gate on qubit 0. We note that \sol{}'s design does not require any additional effort in terms of calculating more reversals or running more circuits. In fact, only one reversed circuit still needs to be run per gate. However, the reversed circuit now contains multiple reversals in order to amplify the error impact of a particular gate on the output of large high-noise circuits.

However, the questions is: are multiple reversals effective and can we determine the number of reversals that are the most effective? We answer this question with experimental evidence. We compare the success of identifying the impact of gates for different reversals across the different algorithms and benchmarks. The success is measured by calculating the Pearson correlation between the TVD obtained by comparing the output of the original circuit and the reversed circuit (the metric we saw above: $TVD\big{(}O_{orig},O_{rev}\big{)}$) and the TVD obtained by comparing the ideal output with the output of the reversed circuit ($TVD\big{(}O_{orig},O_{ideal}\big{)}$). Because getting $O_{ideal}$ requires simulation, it is not practical to calculate as a part of \sol{}'s technique. However, here we use it to validate the success of the technique. \textit{If \ $TVD\big{(}O_{ideal},O_{rev}\big{)}$ is correlated with $TVD\big{(}O_{orig},O_{rev}\big{)}$ across all gates in a circuit, it means that both metrics produce the same relative impact scores for all the gates.} This in turn implies that $O_{orig}$ can be used as a scalable alternative to $O_{ideal}$. So we calculate this correlation for different number of reversals and identify the number for which the correlation is the highest.

Table~\ref{tab:revcorr} shows the results of these correlations across all the algorithms and benchmarks used for evaluating \sol{} for one reversal, three reversals, five reversals, and seven reversals (i.e., seven back-to-back reversed pairs). As the table shows, the correlation values for one reversal are low across different algorithms. In general, the three reversals observe a higher correlation than one reversal and five reversals observe a higher correlation than three reversals. Beyond five reversals, the correlation values either remain steady or drop slightly due to noise. For example, for the 5-qubit Multiply algorithm, the one reversal correlation is 0.76, for three reversals, it is 0.96, and for five and seven reversals, it is 0.99. On the other hand, for the 8-qubit TFIM algorithm, the one reversal correlation is 0.38, the three reversal correlation is 0.53, the five reversal correlation is 0.71, but for seven correlations, it drops down to 0.60. Based on this empirical evidence, \sol{} selected five reversals as it was experimentally confirmed to be effective at identifying the impact of the gates on the final output error. However, we note that \sol{}'s design allows the number of reversals to be configurable and changed if needed. In our experience, the insights and trends generated by \sol{} are consistent for number of reversal beyond three.

As the above results show, \sol{} requires one run per gate in the circuit, which adds up as the number of gates increases. Therefore, next we look at how we can reduce the number of runs required to figure out the impact of gates.

\begin{figure*}[t]
    \centering
    \includegraphics[scale=0.62]{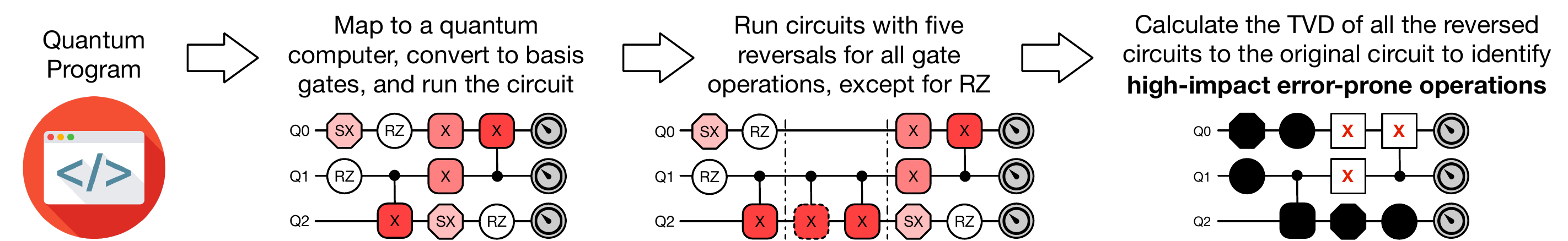}
    \vspace{1mm}
    \hrule
    \vspace{1mm}
    \caption{Summary of the steps used for applying the \sol{} methodology to identify high-impact error-prone quantum gate operations in a quantum program circuit using individual-operation reversals.}
    \label{fig:charter}
    \vspace{-3mm}
\end{figure*}

\subsection{\sol{}: Reducing the Number of Runs}

As \sol{} only takes $g$ circuit runs for $g$ gates operations, it only scales linearly in terms of the number of gate operations: its time complexity is $O(g)$. However, the number of runs can be further reduced if we consider the gate type. Recall that on IBM computers, the basis gates are \cx{}, \rz{}, \sx{}, and \x{}. While \cx{}, \sx{}, and \x{} are physical gates that are implemented on the quantum computers, the \rz{} gate performs a frame-of-reference change virtually and is not implemented on the 
physical quantum computers~\cite{carvalho2021error}. Therefore, these gates have negligible to no impact on the output error of a quantum program (as we will also see in the evaluation section: Sec.~\ref{sec:eval}). As a result, they can be disregarded for the purposes of characterizing the error impact of different gates and their reversed runs can be eliminated from consideration.

Table~\ref{tab:rzcxs} shows the percentage of all quantum gate operations that are the \rz{} operation across different quantum algorithms. The table also shows the percentage of \cx{} gates (the remaining gates are \sx{} or \x{}) and the circuit depth (number of gates in the critical path of the algorithm circuit) for completeness. The table demonstrates that in general, around 20-40\% of the gates in quantum algorithm tend to be \rz{} gates. For example, for algorithms like QFT (7) and TFIM (8), this the percentage of \rz{} gates can be as high as 42\%. \textit{This directly implies that the number of runs required by \sol{} can be reduced by 20-40\% in practice depending on the algorithm,} even though the complexity remains $O(g)$ in the worst case.

Next, we summarize the overall approach of \sol{}.

\subsection{\sol{}: Putting it All Together}

In Fig.~\ref{fig:charter} we summarize the different steps of \sol{}. \sol{} takes the quantum program and maps it on to the quantum hardware and converts it into the hardware basis gates. Then for each of the gates, it generates reversed circuits, each with their own pair of the reversed gate + the original gate applied right next to the original gate. The pair application is repeated five times over to generate the best success in terms of identifying the high-impact gates in the circuit. This is done for all gates except for the \rz{} gates as these gates are virtual and have negligible error impact. The next step is to run the original circuit and all the reversal circuits on the quantum computers. Once the runs are completed, the TVDs between the original circuit and the reversed circuits can be calculated and the high-impact gates can be determined by identifying the reversed circuits with high TVDs.

In the following section, we evaluate the effectiveness of \sol{} in identifying such high-impact gates and derive some surprising and interesting insights in the process.

\begin{figure*}[t]
    \centering
    \subfloat[QFT (3) Circuit for Output Hamming Weight One]{\includegraphics[scale=0.355]{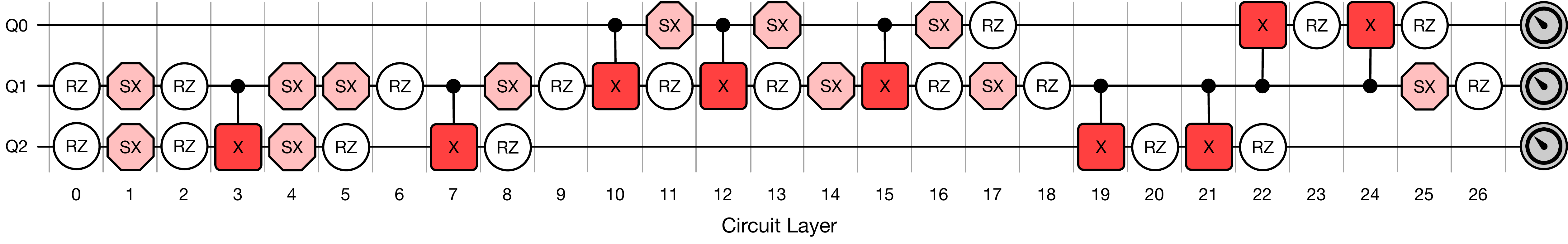}}\\
    \subfloat[QFT (3) Output Hamming Weight Zero]{\includegraphics[scale=0.44]{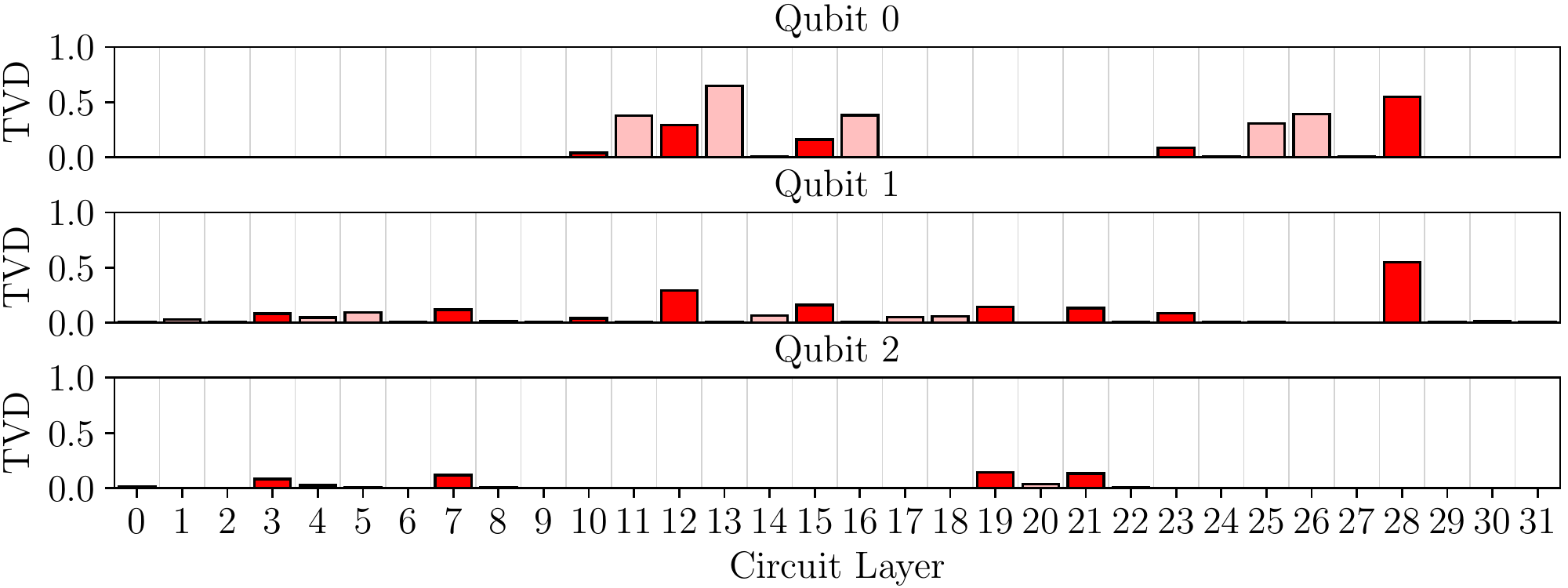}}\hfill
    \subfloat[QFT (3) Output Hamming Weight One]{\includegraphics[scale=0.44]{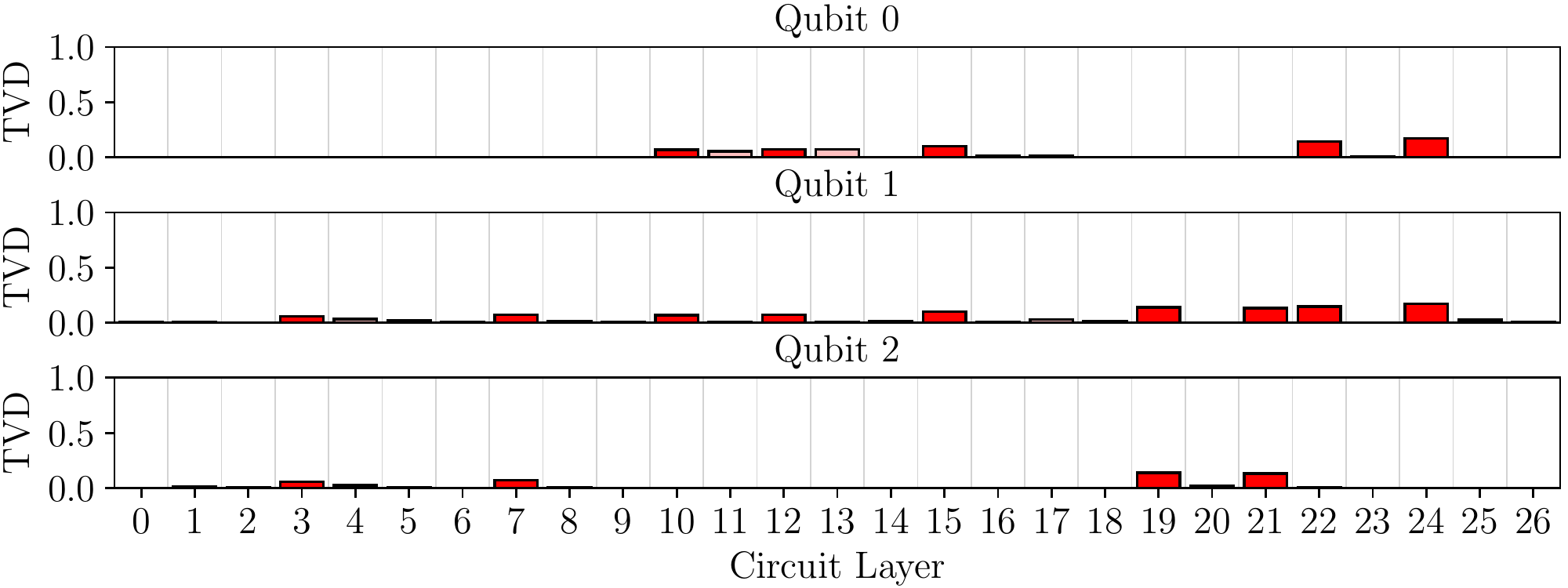}}\\
    \subfloat[QFT (3) Output Hamming Weight Two]{\includegraphics[scale=0.44]{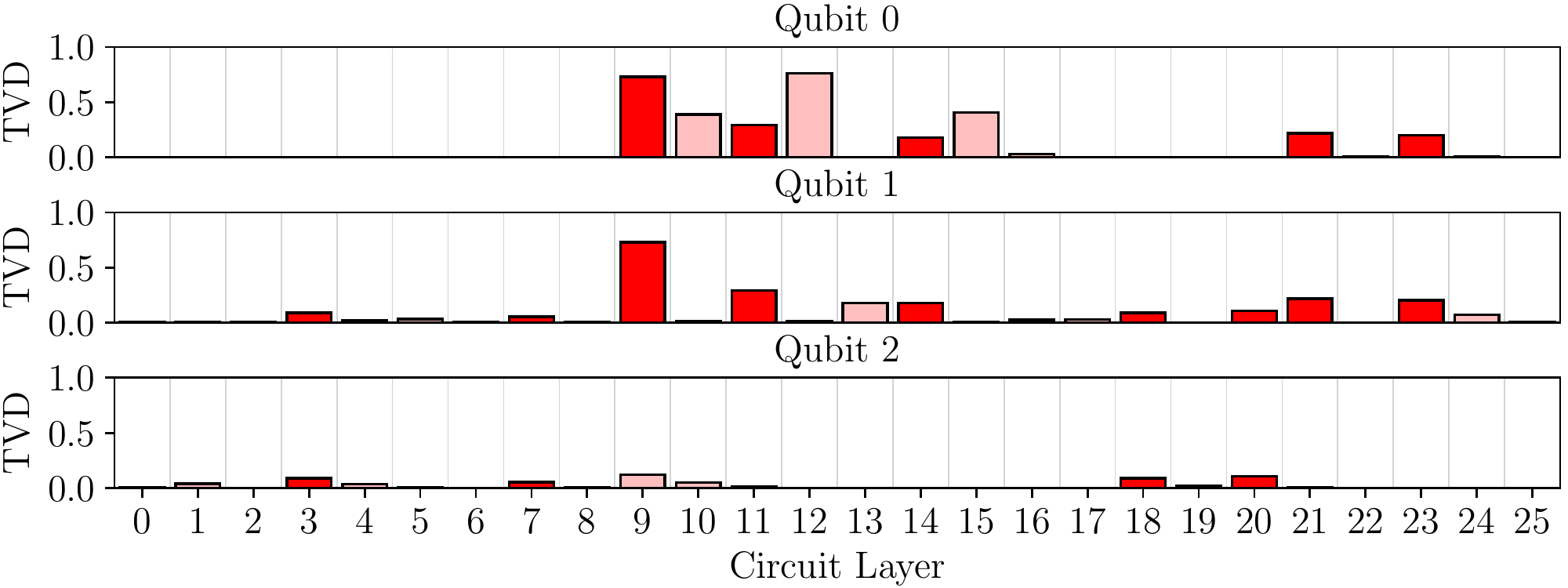}}\hfill
    \subfloat[QFT (3) Output Hamming Weight Three]{\includegraphics[scale=0.44]{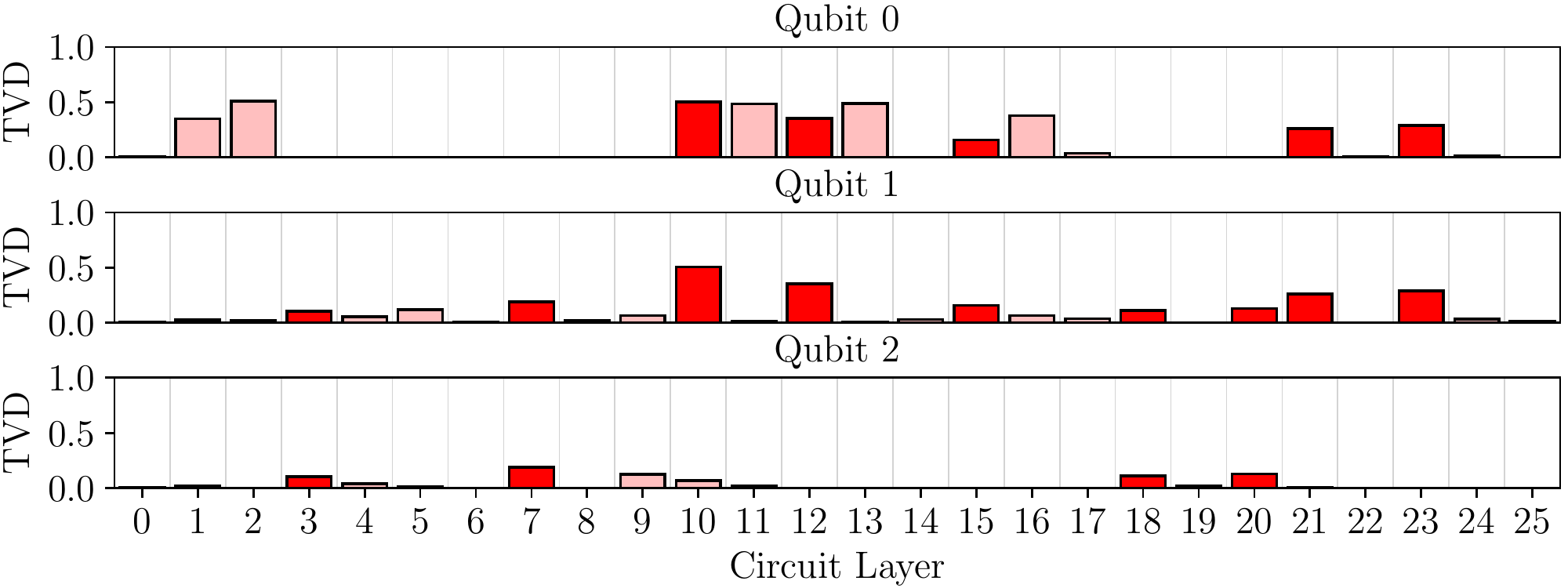}}
    \vspace{1mm}
    \hrule
    \vspace{1mm}
    \caption{(a) The circuit structure of the three-qubit QFT algorithm. TVD of the three-qubit QFT circuits between the reversed circuits and the original circuit when it is run with input such that the output has a hamming weight of (b) 0, (c) 1, (d) 2, and (e), 3. The results show that depending on the input, the impact of the different gates in the circuit can vary.}
    \label{fig:qft}
    \vspace{-3mm}
\end{figure*}
\section{\sol{}: Evaluation, Analysis, and Insights}
\label{sec:eval}

This section is organized as following: we perform different types of analyses and present a summary of observations and insights for each of those. These analyses are performed for all the algorithms and benchmarks used for evaluation and general trends are highlighted. We also note diverging patterns when they are significant, and we provide strategies to mitigate the error impact of high-impact gates.

Before we present these general results, we first perform a detailed case study using the three-qubit QFT algorithm, in order to better appreciate the insights derived from the combined trends presented later. This case study includes a breakdown of the error impact of individual gates in the QFT algorithm and an examination of how this error impact varies with the input provided to the algorithm.



\vspace{3mm}

\noindent\textbf{QFT(3) Case Study and the Influence of the Input.} Fig.~\ref{fig:qft}(a) shows the circuit structure of the three-qubit QFT algorithm. We see that the algorithm has 39 operations in total: 9 \cx{}, 18 \rz{}, 12 \sx{}. So we run 39 reversed circuit runs + 1 original circuit run $=$ 40 runs (combined runtime of $<0.5$ hours on the IBM computers). We also see that the 39 operations are organized in 27 layers of parallel executions. Note that for this algorithm we also run the reversed runs for the \rz{} gates to demonstrate that the \rz{} gates have negligible impact (but, in general we do not need to execute the runs corresponding to the \rz{} gates). We calculate the 39 TVDs of the output of the reversed runs to the output of the original run.

Fig.~\ref{fig:qft}(b) shows these TVDs. For easier interpretation, the figure has been structured into three sub-figures that show the results for the three qubits (qubit 0 at the top, 1 in the middle, and 2 at the bottom). The x-axis shows the layer at which the operation took place, corresponding to the layers indicated at the bottom in Fig.~\ref{fig:qft}(a). The bars show the TVD for the individual gates on different qubits. The positions of the bars correspond to the positions of the gates in Fig.~\ref{fig:qft}(a) (the bar colors also correspond to the operation colors). For two-qubit \cx{} gates, the bar is drawn twice on both the qubits. For example, the bar corresponding to the \cx{} gate on qubits 0 and 1 in layer 12 is drawn twice: at layer 12 on qubit 0 and at layer 12 on qubit 1. As the TVD ranges between 0 and 1, the y-axis of all three figures is from 0 to 1.

\textit{The first observation we make is that the \rz{} gates have negligible impact.} The bars corresponding to the \rz{} gates are not visible due to the TVD being very low. For example, there are two \rz{} gates in layer 0 and layer 2 on qubits 1 and 2 in Fig.~\ref{fig:qft}(a). However, when we look at layers 0 and 2 in Fig.~\ref{fig:qft}(b), the bars are not visible due to their low impact. The lack of a visible bar can be verified for all \rz{} gates in the circuit. This demonstrates that the runs corresponding to the \rz{} gates can be eliminated from the characterization process. In the case of the three-qubit QFT algorithm, this means that the number of runs can be reduced from 40 to $40-18=22$, which is a reduction of 45\% in terms of the execution overhead.

\textit{Next, we identify the gates with the highest impact on the output error.} In Fig.~\ref{fig:qft}(b), we ran the three-qubit QFT circuit with input specified to ensure that the output has a hamming weight of zero (i.e., ``000''). Fig.~\ref{fig:qft}(c), (d), and (e) shows what happens for other inputs. In Fig.~\ref{fig:qft}(c), we ran the three-qubit QFT circuit for a hamming weight of one (i.e., ``001'' or ``010'' or ``100''). Similarly, Fig.~\ref{fig:qft}(d) shows the results for hamming weight of two (i.e., ``011'' or ``101'' or ``110''), and Fig.~\ref{fig:qft}(e) shows the results for hamming weight of three (i.e., ``111''). \textit{Note that the different inputs may change the circuit structure at the beginning of circuit, but the main section of QFT (starting at layer 3 in Fig.~\ref{fig:qft}(a)) remains the same.}

\textit{Changing the input for the QFT algorithm has a considerable effect on the impact that different gates have on the overall output error of the circuit.} For example in Fig.~\ref{fig:qft}(b), the gates with the highest impact include the \cx{} gate on qubits 0 and 1 in layer 28. However, the \sx{} gate on qubit 0 in layer 13 has a higher impact. In fact, in general, \sx{} gates in the middle and end sections tend to have a high impact in Fig.~\ref{fig:qft}(b). \textit{It is notable that in terms of the physical implementation, the same \sx{} gate is being applied to qubit 0 in layers 11, 13, 16, 25, and 26. However depending on its position, it has a varying impact on the output error.} On the other hand, in Fig.~\ref{fig:qft}(c), the tallest TVD bars exist corresponding to four \cx{} gates: (I) qubit 1 and 2 in layer 19, (II) qubits 1 and 2 in layer 21, (III) qubits 0 and 1 in layer 22, and (IV) qubits 0 and 1 in layer 24. \sx{} gates are not high-impact for this input. In contrast, in Fig.~\ref{fig:qft}(d), the \cx{} and \sx{} gates in the middle section of the circuit (layers 9 to 15) on qubits 0 and 1 have the highest impact compared to the \cx{} gates at the end, and in Fig.~\ref{fig:qft}(e), the \cx{} and \sx{} gates all across the length of the circuit (layers 1-2, 10-16, and 21-23) on qubits 0 and 1 have a high impact on the overall output error.

\vspace{3mm}

\noindent\textbf{
Use case:\sol{} for Discovering High-Impact Inputs.} \textit{To figure out which input has the highest impact on the output error, \sol{} is leveraged as a technique to apply reversals to all input-related gates together (instead of one gate a time) to estimate their combined impact.} All input related gates are bunched together and reversed within the same circuit, instead of in different circuits. Applying this technique produces TVDs of the outputs of the reversed circuits to the output of the original circuit as following: TVD $=$ 0.06 for hamming weight zero input (Fig.~\ref{fig:qft}(b)), 0.02 for hamming weight one input (Fig.~\ref{fig:qft}(c)), 0.06 for hamming weight two input Fig.~\ref{fig:qft}(d), and 0.07 for hamming weight three input (Fig.~\ref{fig:qft}(e)). \sol{} reveals that the input in Fig.~\ref{fig:qft}(e) (hamming weight of three) has the highest impact, which is also evidenced by the fact that it is the only input with high-impact gates in layers 1-2.

\vspace{2mm}

\begin{mybox}{red}{red}
\textbf{Observation I.} Previous works that focus on reducing the output error on noisy quantum computers using compiler optimizations focus on high-level optimizations based on simplistic assumptions~\cite{wilson2020just,patel2020ureqa,murali2019noise,patel2020veritas}. As an instance, previous works have focused on optimizing the mapping to hardware based on error information of the gates obtained during computer calibration in isolation or via micro-benchmarking, that is, one number is used per physical gate type~\cite{wilson2020just,tannu2019ensemble,tannu2019not}. However, \textbf{\sol{} has shown that the same physical gate can have very different impact based on its position within the circuit -- this requires critical consideration for compiler optimizations which is currently lacking in existing quantum compiler solutions.}
\end{mybox}

\vspace{1mm}

\begin{mybox}{red}{red}
\textbf{Observation II.} Using \sol{}, we also showed that depending on the changes in the input, which changes the gate structure at the beginning of the circuit, the impact of the different gates will vary even if the rest of the circuit structure is the same. This means optimizations that are based purely based on circuit structure~\cite{li2022paulihedral,bhattacharjee2019muqut,davis2020towards} (e.g., gate deletion and synthesis~\cite{tan2020optimality,patel2022quest,amy2013meet,wilson2021empirical} and layout mapping and routing~\cite{zhang2021time,zulehner2018efficient,li2019tackling}) may be sub-optimal due to the lack of input-awareness. \textbf{An effective demonstrated use case of \sol{}'s includes performing a combined reversal on all input-related gates to identify the inputs with the highest impact on the output error.}
\end{mybox}

\vspace{3mm}

\noindent\textbf{The Effect of Gate Position Along the Length of the Circuit.} In the three-qubit QFT examples for different inputs, we saw that gates in any layer position along the circuit length can be high-impact gates (e.g., Fig.~\ref{fig:qft}(e)). However, conventional wisdom suggests that error impacts are especially worse near the end of the circuits due to the qubit state losing coherence. Recall that over time, the qubit state tends to lose its phase and amplitude and decay to the ground (i.e., the $\ket{0}$) state~\cite{mavadia2017prediction,PhysRevA.58.2733}. So this effect gets particularly worse at the end layers of a quantum circuit execution.

While previously there did not exist a way to test the above conventional wisdom, now \sol{} enables us to determine if this conventional wisdom is true. To perform this test, we correlate the impact of the gates (i.e., the TVD between the reversed circuit corresponding the gate and the original circuit) with the layer in which it is positioned. If the correlation is high, it suggests that as the layer index increases, the gate impact also increases, that is, it suggests that the assumption is true. For example, say we have gates $G = \{g_1, g_2, g_3, g_4, g_4\}$ with TVDs $T = [0.1, 0.2, 0.3, 0.32, 0.4]$; the gates are positioned in layers $L = [1, 2, 3, 3, 4]$. If we correlate the TVD vector $T$ with the layers vector $L$, we would observe a high correlation. If instead, $T = [0.2, 0.1, 0.1, 0.2, 0.1]$, we would observe a low correlation.

\begin{table}[t]
    \centering
    \caption{Pearson correlation between the error impact of a gate and its position within the program (the layer number). The correlation is low or insignificant for most algorithms.}
    \vspace{-1mm}
    \begin{tabular}{|>{\columncolor[gray]{0.85}}p{13mm}|>{\columncolor[RGB]{255, 200, 200}}p{7mm}|p{9mm}||>{\columncolor[gray]{0.85}}p{15.5mm}|>{\columncolor[RGB]{255, 200, 200}}p{7mm}|p{9mm}|}
        \hline
        \textbf{Algorithm} & \textbf{Corr.} & \textbf{$p$-value} & \textbf{Algorithm} & \textbf{Corr.} & \textbf{$p$-value}\\
        \hline
        \hline
        HLF (5) & -0.04 & 0.79 & QAOA (5) & 0.43 & 2$e-$7\\
        \hline
        HLF (10) & 0.14 & 0.05 & QAOA (10) & 0.29 & 9$e-$9\\
        \hline
        QFT (3) & 0.17 & 0.27 & VQE (4) & 0.21 & 1$e-$5\\
        \hline
        QFT (7) & $-$0.66 & 4$e-$37 & Heisen. (4) & 0.27 & 2$e-$10\\
        \hline
        TFIM (4) & 0.12 & 0.20 & Adder (4) & $-$0.02 & 0.84 \\
        \hline
        TFIM (8) & 0.33 & 2$e-$15 & Adder (9) & 0.05 & 0.78 \\
        \hline
        TFIM (16) & 0.26 & 1$e-$9 & Multiply (5) & 0.1 & 0.36 \\
        \hline
        XY (4) & -0.14 & 0.18 & Multiply (10) & 0.58 & 4$e-$60 \\
        \hline
        XY (8) & 0.42 & 1$e-$22 & - & - & - \\
        \hline
    \end{tabular}
    \label{tab:lencorr}
    \vspace{0mm}
\end{table}

\sol{} performs this correlation for all evaluated algorithms and the results are reported in Table~\ref{tab:lencorr}. The table shows the Pearson correlation numbers along with the corresponding $p$-values. As the table shows, for most algorithms (e.g., HLF (5), QFT (3), Adder (9), Multiply (5), TFIM (4), and XY (4)), the correlation numbers are not high and the $p$-value numbers are high ($>0.01$), which indicates that no strong correlation exists between the impact of a gate and the layer in which it is located. The highest correlation among all algorithms of 0.58 can be seen for the Multiply (10) algorithm. On the other side of the spectrum, the QFT (7) algorithm has a correlation of $-$0.66. The negative correlation indicates that most of its high impact gates are located in the front section of the circuit.

\textit{Overall, the results highlight that for most quantum algorithms, the high-impact gates can be distributed somewhat evenly across the entire length of the circuit.} While some algorithms may have a concentration of high-impact gates at the back of the circuit, some others may have these gates concentrated at the beginning of the circuit. The behavior varies from algorithm to algorithm with circuit structure.

\vspace{3mm}

\noindent\textbf{Analysis of Multiple Architectures using VQE.} To demonstrate that \sol{}'s analysis are applicable to different quantum computing architectures and topologies, we run the VQE circuit on \texttt{ibmq\_quadalupe}, along with the default execution on \texttt{ibm\_lagos}. As the VQE circuit is a four-qubit circuit, we choose the first four qubits of the two computers for execution to make use of different topologies; Fig.~\ref{fig:comps} shows qubits 0, 1, 2, and 3 form a T-shape on \texttt{ibm\_lagos} and a line on \texttt{ibmq\_quadalupe}. Due to this difference in topologies, when VQE is mapped to \texttt{ibmq\_quadalupe}, it has 135 \rz{} gates -- as opposed to the the default of 172 on \texttt{ibm\_lagos} -- and 74 \cx{} gates -- as opposed to 132 on \texttt{ibm\_lagos}. However, in spite of these differences, the correlation between the error impact of a gate and its position within the program is 0.41 ($p-$value: $1e-13$) on \texttt{ibmq\_quadalupe}; it is 0.21 on \texttt{ibm\_lagos}. The low correlation on both computers shows that \sol{}'s analysis is applicable to different computer architectures and topologies.

\vspace{2mm}

\begin{mybox}{red}{red}
\textbf{Observation III.} Due to the property of NISQ qubits of losing their coherence over time, it has been assumed that a large portion of the output error is accumulated near the end of the circuit based on the study of decoherence effects in isolation~\cite{mavadia2017prediction,PhysRevA.58.2733}. In fact, some previous works have proposed software-level compiler optimizations that are especially catered toward the end or the beginning of the quantum circuit~\cite{liu2021relaxed,shi2019optimized}.

\hspace{2mm} \textbf{\sol{} demonstrate that these conventional assumptions do not hold true for many real-world quantum algorithms with complex circuit structures that face the combined effect of all different types of hardware noises, when they are executed on different quantum computer architectures and topologies}. Depending on the circuit properties of the algorithm, the high-impact gates can be evenly distributed, they can be bundled in the front section of the circuit, or they can be bundled in the back section.
\end{mybox}


\begin{table}[t]
    \centering
    \caption{Percentage of all program qubits that appear in top 5\%, 10\%, 25\% and 50\% of high-error-impact gates. High-impact gates can exist across different qubits for all algorithms.}
    \vspace{-1mm}
    \begin{tabular}{|>{\columncolor[gray]{0.85}}p{20mm}|p{10mm}|p{12mm}|>{\columncolor[RGB]{255, 200, 200}}p{12mm}|p{12mm}|}
        \hline
        \textbf{Algorithm} & \textbf{Top 5\%} & \textbf{Top 10\%} & \textbf{Top 25\%} & \textbf{Top 50\%} \\
        \hline
        \hline
        HLF (5) & 40\% & 40\% & 60\% & 100\% \\
        \hline
        HLF (10) & 70\% & 100\% & 100\% & 100\% \\
        \hline
        QFT (3) & 67\% & 67\% & 100\% & 100\% \\
        \hline
        QFT (7) & 57\% & 71\% & 86\% & 100\% \\
        \hline
        Adder (4) & 100\% & 100\% & 100\% & 100\% \\
        \hline
        Adder (9) & 78\% & 100\% & 100\% & 100\% \\
        \hline
        Multiply (5) & 40\% & 60\% & 100\% & 100\% \\
        \hline
        Multiply (10) & 90\% & 100\% & 100\% & 100\% \\
        \hline
        QAOA (5) & 40\% & 60\% & 60\% & 100\% \\
        \hline
        QAOA (10) & 90\% & 90\% & 100\% & 100\% \\
        \hline
        VQE (4) & 100\% & 100\% & 100\% & 100\% \\
        \hline
        Heisenberg (4) & 100\% & 100\% & 100\% & 100\% \\
        \hline
        TFIM (4) & 75\% & 100\% & 100\% & 100\% \\
        \hline
        TFIM (8) & 88\% & 100\% & 100\% & 100\% \\
        \hline
        TFIM (16) & 94\% & 100\% & 100\% & 100\% \\
        \hline
        XY (4) & 50\% & 50\% & 100\% & 100\% \\
        \hline
        XY (8) & 100\% & 100\% & 100\% & 100\% \\
        \hline
    \end{tabular}
    \label{tab:qubcorr}
    \vspace{0mm}
\end{table}

\vspace{3mm}

\noindent\textbf{The Impact of Individual Qubit's Error Rates.} In the QFT (3) results in Fig.~\ref{fig:qft}, we observed that most of the high-impact gates were situated on qubit 0, then there were a few on qubit 1, and there were no high-impact gates on qubit 2. This example is a good representation of the common way of performing noise characterization of quantum computers. Many works focus on characterizing the noise effects at the level of individual qubits or at the level of specific types of operations performed on individual qubits~\cite{erhard2019characterizing,proctor2019direct,patel2020experimental,liu2020reliability,butko2019understanding,wright2019benchmarking}. For example, previous works have relied on data obtained during computer calibration to estimate the error rates of qubits to perform noise-aware compilation and reduce the gate count on high-error qubits~\cite{wilson2020just,tannu2019not,patel2020ureqa,murali2019noise,patel2020veritas}.



\sol{}'s novel methodology enables us to investigate if this broad classification of qubits as being high-error or low-error based on simplistic noise models or isolation noise analysis holds true when we look at the impact of individual gates. In other words, we want to study if all the high-impact gates are concentrated on a few qubits (which would encourage us to broadly classify those qubits as high-error) or if these high-impact gates are distributed across the different qubits that are involved in the quantum circuit.

First, we note that, unlike the case of studying the concentration of high-impact gates across the length of the circuit, we cannot use correlation to perform this check as the actual qubit ids are not of relative significance. For example, layer 1 is after layer 0 for a correlative study but qubit ids have no relative relationship. Therefore, instead, in Table~\ref{tab:qubcorr}, we show the percentage of all qubits in a circuit that are involved in the top 5\% (i.e., gates with the top 5\% highest TVD between the corresponding reversed circuits and the original circuit), 10\%, 25\%, and 50\% of high-impact gates. Note that we do not use just one threshold to define high impact, as this threshold may vary from algorithm to algorithm and use-case to use-case.

As the results show, for most of the algorithms, a majority of the qubits are involved in top 5\% of the high-impact gates and a 100\% of the qubits are involved in the top 25\% of high impact gates. Even for relatively larger algorithms like TFIM (16), 15/16 qubits (94\%) are involved in the top 5\% of high impact gates. If we look at the top 50\% column, we see that a 100\% of the qubits are involved in high-impact gates across all of the evaluated algorithms. \textit{These results suggest that most qubits can have gates that have a high-impact on the output error and it is preferable to identify the error of specific gates, than to broadly classify qubits as being high error or low error.}

\vspace{2mm}

\begin{mybox}{red}{red}
\textbf{Observation IV.} General characterization and calibration work in the NISQ computing regime tries to broadly characterize the error rates of individual qubits on a quantum computer~\cite{erhard2019characterizing,proctor2019direct,patel2020experimental,liu2020reliability,butko2019understanding,wright2019benchmarking}. Compiler-level optimization works then try to use these board classifications to perform noise-aware mapping and routing of quantum algorithms to the quantum computers~\cite{wilson2020just,patel2020ureqa,tannu2019not,murali2019noise,patel2020veritas}.

\hspace{2mm} \textit{\sol{}'s unique methodology has helped us reveal that the gates that have a high-impact on the output error of a quantum program can exist at different places within the circuit and on any qubit.} In addition, \textbf{we have demonstrated that the top 25\% of high-impact operations are distributed across all the qubits for most quantum algorithms. \sol{} shows that simply classifying qubits based on their error rate and focusing software optimizations around this classification may not be sufficient and complete -- criticality of gate operations is equally important and they can occur on low-error qubits too.}
\end{mybox}

\vspace{3mm}

\noindent\textbf{The Role of One-Qubit Gates in Output Error.} In the QFT (3) results in Fig.~\ref{fig:qft}, we observed that both \cx{} and \sx{} gates can be high-impact gates, depending on the input to the program. However, due to the fact that \cx{} gates tend to have an order of magnitude higher error rate than \sx{} or \x{} gates when measured in isolation~\cite{patel2020veritas,patel2020experimental}, many software-level compiler works have specifically focused on minimizing the number of \cx{} gates in the circuits. This has been in the form of directing trying to reduce the \cx{} gates using mathematical equivalencies based on the commutativity and unitary properties of gates within the circuit~\cite{smith2019quantum,zulehner2019compiling}, and in the form of trying to reduce the number of \texttt{SWAP} operations, each of which is implemented using three \cx{} gates~\cite{wille2019mapping,ash2019qure}. 


While these techniques are effective at reducing the number of \cx{} gates, they also have a significant compilation time overhead. \sol{} can be used to study if these techniques are worth the overhead considering that they only focus on \cx{} gates, which are assumed to be high-impact due to their relatively higher isolation error rates. We also saw for the QFT (3) example, that some \sx{} gates can have a much higher impact than \cx{} gates. To investigate if this is the case across different algorithms, in Table~\ref{tab:sx_x}, we present the percentage of one-qubit \sx{} and \x{} gates in the circuit that have a greater error impact than the \cx{} gate with the least error impact compared to other \cx{} gates in the circuit. We compare to the \cx{} gate with the least error impact as all \cx{} gates are targeted for compiler optimizations regardless of their specific error impact. So the one-qubit \sx{} or \x{} gates that have a higher impact than that \cx{} gate can be judged to have a high impact overall -- this can be caused by potentially higher crosstalk effects and neighboring gate operations around the qubits where \sx{} and \x{} gates are being performed.

\begin{table}[t]
    \centering
    \caption{Number and percentage of one-qubit \sx{} and \x{} gates that have a greater error impact than the \cx{} gate with the least error impact (compared to other \cx{} gates). A large portion of one-qubit gates have a higher impact than \cx{} gates.}
    \vspace{-1mm}
    \begin{tabular}{|>{\columncolor[gray]{0.85}}p{13mm}|p{7mm}|>{\columncolor[RGB]{255, 200, 200}}p{8mm}||>{\columncolor[gray]{0.85}}p{17mm}|p{7mm}|>{\columncolor[RGB]{255, 200, 200}}p{8mm}|}
        \hline
        \textbf{Algorithm} & \textbf{Num. \sx{}+\x{}} & \textbf{\%\newline \sx{}+\x{}} & \textbf{Algorithm} & \textbf{Num. \sx{}+\x{}} & \textbf{\%\newline \sx{}+\x{}}\\
        \hline
        \hline
        HLF (5) & 7 & 70\% & QAOA (5) & 22 & 71\%\\
        \hline
        HLF (10) & 45 & 92\% & QAOA (10) & 58 & 89\%\\
        \hline
        QFT (3) & 9 & 56\% & VQE (4) & 119 & 98\%\\
        \hline
        QFT (7) & 78 & 98\% & Heisenberg (4) & 141 & 96\%\\
        \hline
        TFIM (4) & 30 & 83\% & Adder (4) & 20 & 74\%\\
        \hline
        TFIM (8) & 179 & 95\% & Adder (9) & 35 & 78\%\\
        \hline
        TFIM (16) & 772 & 98\% & Multiply (5) & 20 & 80\%\\
        \hline
        XY (4) & 21 & 75\% & Multiply (10) & 117 & 100\%\\
        \hline
        XY (8) & 158 & 98\% & - & - & - \\
        \hline
    \end{tabular}
    \label{tab:sx_x}
    \vspace{0mm}
\end{table}

The results in the table show that across different algorithms, 50\% to over 90\% of all \sx{} + \x{} gates tend to have a higher impact than the \cx{} gate with the least error impact in the circuit. For example, for the XY (8) algorithm, 98\% of the \sx{} + \x{} gates have a higher error impact. \textit{These results highlight the need to focus on impact-aware compiler optimizations for gates instead of type-aware.}

\vspace{3mm}

\noindent\textbf{\sol{}'s Mitigation Strategy. This above observation about high-impact one-qubit gates is also evidenced by the fact that the \sx{} gates on qubit 0 in layers 1 and 2 have some of the highest impact in Fig.~\ref{fig:qft}(e). This is because there are parallel \sx{} gates on qubits 2 and 3 in layers 1 and 2 for this input, which causes increased crosstalk effect on qubit 0. \textit{As a mitigation strategy, we inserted barriers to make sure the gates in these two layers are executed serially instead of in parallel and observed the output error (TVD of the observed output to the ideal output) drop from 0.19 to 0.12 (i.e., 7\% points reduction).} Note that all operations in a circuit cannot be serialized as this would increase circuit depth and decoherence effects. However, this strategy is effective when used selectively for the highest-impact gates.}

\vspace{2mm}

\begin{mybox}{red}{red}
\textbf{Observation V.} Due to the fact that \cx{} gates have an order of magnitude higher error rate when the error rate is measured in isolation~\cite{patel2020veritas,patel2020experimental}, it is often assumed that these gates also have the highest error impact in quantum circuits when run on NISQ computers~\cite{ash2019qure,tan2020optimality,smith2019quantum}. Previous works have focused on minimizing the impact of \cx{} gates by trying to reduce their number in quantum circuits~\cite{zulehner2019compiling,wille2019mapping,patel2022quest,amy2013meet,wilson2021empirical}.

\hspace{2mm} In contrast, we have leveraged \sol{} to demonstrate an interesting insight: one-qubit \sx{} and \x{} gates also have a high impact on the output error, and in some cases this impact is higher than that of \cx{} gates. In fact, across different quantum algorithms, 50\%-98\% of all \sx{} + \x{} gates have a higher error impact than the \cx{} gate with the least error impact in the algorithm circuit. \textbf{\sol{} serves as a useful utility for the programmer to not rely on generalized assumptions and determine gate-specific error impact. As show by \sol{}, identification and selective serialization of gates can act as an effective mitigate strategy for reducing the impact of high-impact gates.}
\end{mybox}
\section{Work Related to \sol{}}
\label{sec:work}

Previous works relevant to \sol{} are as following:

\vspace{3mm}

\noindent\textbf{Noise Characterization.} Previous works have focused on characterizing different aspects of the noise on NISQ computers, from examining a specific type of noise such as decoherence effects~\cite{mavadia2017prediction,PhysRevA.58.2733}, to benchmarking and micro-benchmarking different gates and algorithms on quantum computers to figure out the kinds of errors they experience~\cite{patel2020experimental,liu2020reliability,butko2019understanding,erhard2019characterizing,proctor2019direct,wright2019benchmarking}. For example, Erhard et al.~\cite{erhard2019characterizing} perform cycle benchmarking on quantum computers to figure out how their noise characteristics manifest in different algorithms. However, \textit{none of these works characterize the impact that a specific gate has on the output error of a quantum program. \sol{} makes a novel contribution in that space.}

\vspace{3mm}

\noindent\textbf{Program Debugging.} Previous works have introduced debugging techniques for quantum code~\cite{huang2019statistical,liu2020quantum,liu2021systematic,yuan2022twist,haner2021distributed}. Some of these works propose the use of statistical assertions to debug the functionality of the circuits~\cite{huang2019statistical,liu2020quantum,liu2021systematic}, while others propose new programming interfaces and frameworks for semantics-based debugging~\cite{yuan2022twist,haner2021distributed}. A concurrent work~\cite{calderon2022quantum} shares the goal with \sol{} of debugging and sensitivity analysis of quantum circuit implementations leveraging the idea of gate reversal. This work can also be used to locate bugs at different levels of a quantum-program specification. We note that \sol{} was submitted to SC for peer review on April 1, 2022 (the submission deadline for full technical paper for SC), prior to the appearance of the arXiv article (the arXiv article was first uploaded on April 12, 2022~\cite{calderon2022quantum}). But, the arXiv article was brought to our attention after \sol{} was published online in mid-November, 2022 as a part of the SC proceedings. Therefore, we are uploading this version to acknowledge and credit the authors of the arXiV article for their idea and contributions to the field. The concurrency of these ideas further demonstrates the potential of reversing gates and circuits for various purposes. \textit{\sol{} can further enhance the capability of these works by enabling gate-specific debugging of errors.}

\vspace{3mm} 

\noindent\textbf{Compiler Optimizations.} Different types of software-based compiler optimizations have been proposed to reduce the impact of noise when quantum algorithms are run on quantum computers, including works that mitigate the effect of crosstalk~\cite{murali2020software,xie2022suppressing}, perform noise-optimal compilation using isolation noise characteristics~\cite{wilson2020just,patel2020ureqa,tannu2019not,murali2019noise,patel2020veritas}, synthesize circuits to reduce their length
~\cite{tan2020optimality,patel2022quest,amy2013meet,wilson2021empirical}, perform layout-aware routing and mapping~\cite{zhang2021time,zulehner2018efficient,li2019tackling}, leverage optimizations targeting specific circuit regions (e.g., end of the circuit)~\cite{liu2021relaxed,shi2019optimized}, reduce the number of \cx{} gate operations~\cite{smith2019quantum,zulehner2019compiling,wille2019mapping,ash2019qure,patel2022quest}, perform optimizations without considering the input~\cite{li2022paulihedral,bhattacharjee2019muqut,tannu2019ensemble,davis2020towards}, and even use the reversibility property at the circuit level to improve the output error (although, it requires ideal simulations on classical computers)~\cite{patel2021qraft}. \textit{While none of these works provide a methodology like \sol{}'s to help programmers learn about the impact that specific gates have on the output of their program, many of them can benefit from this functionality to further improve the compiler optimizations that are applied. }

\section{\sol{}: Concluding Remarks}
\label{sec:concl}

In this paper, we proposed \sol{}, a simple and effective technique that \textit{leverages reversibility property of quantum circuits to identify high-impact error-prone gate operations in quantum programs}. \sol{} demonstrates the usability of quantum operation reversibility for identifying most critical quantum operations -- extending the application of quantum operation reversibility beyond estimating and improving answer fidelity of quantum programs~\cite{patel2021qraft}.

By applying \sol{} to a variety of quantum algorithms and benchmarks with different characteristics, we highlighted several surprisingly and counter-intuitive trends, such as the observation that high-impact error-prone gates can be located anywhere within a quantum program. \sol{}'s open-source implementation and dataset is available to future innovations in compiler optimization and algorithm debugging techniques. 

\section*{Acknowledgement}

We thank the anonymous reviewers for their constructive feedback. This work was also supported by Northeastern University, NSF Award  1910601 and 2144540. IBM Q was also used for this work. The views expressed are those of the authors and do not reflect the official policy or position of IBM or the IBM Q team.

\balance
\bibliographystyle{plain}
\bibliography{main}

\end{document}